\documentclass[twocolumn,final,aps,prb,showpacs,superscriptaddress,amsmath,amssymb,amsfonts,floatfix]{revtex4-1}

\usepackage{graphicx}
\usepackage{multirow}
\usepackage{epsfig}
\usepackage{url}
\usepackage{color}
\usepackage{grffile}

\usepackage[colorlinks,breaklinks,bookmarks=true,citecolor=blue,linkcolor=blue,urlcolor=blue]{hyperref}
\usepackage{color}
\usepackage{tabularx}
\usepackage{sidecap}
\usepackage{soul}
\usepackage{float}

\begin{document}

\title{Non-Fermi liquid behavior of scattering rate
in three-orbital Emery model}
\author{Tianzong Mao}
\affiliation{Institute of Theoretical and Applied Physics, Jiangsu Key Laboratory of Thin Films, School of Physical Science and Technology, Soochow University, Suzhou 215006, China}

\author{Mi Jiang}
\affiliation{Institute of Theoretical and Applied Physics, Jiangsu Key Laboratory of Thin Films, School of Physical Science and Technology, Soochow University, Suzhou 215006, China}


\begin{abstract}
Motivated by the recent findings on the $T$-linear electronic scattering rate in the two-dimensional Hubbard model, we have investigated the three-orbital Emery model and its temperature-dependent electronic and quasiparticle scattering rates by adopting dynamical cluster quantum Monte Carlo simulations. By focusing on two characteristic site energies $\epsilon_p$ of O-2$p$ orbital relevant to cuprates and nickelates separately, our exploration discovered that, for $\epsilon_p=3.24$ relevant to cuprates, the scattering rate can exhibit a linear-$T$ dependence at low temperature for a range of intermediate densities. In contrast, for larger $\epsilon_p=6.0$ presumably relevant to nickelates, a wide range of densities support a downturn of the scattering rate below the temperature scale $T\sim0.1$ with possibly two consecutive nearly linear-$T$ regimes connected via a smooth crossover around $T\sim0.1$. 
Furthermore, the temperature dependent quasiparticle scattering rate generically departs from the unity slope as predicted by the Planckian dissipation theory. 
Our presented work provides valuable insights on the extensively studied three-orbital Emery model, particularly on the quantitative examination of non-Fermi liquid features of scattering rates. 
\end{abstract}

\maketitle

\section{Introduction}
As a central topic in condensed matter physics, non-Fermi liquid (NFL) phenomenology is ubiquitous and has been extensively studied in a wide variety of materials~\cite{Rev1,Rev2,Rev3,Rev4,1,2,3,4,5}. One notable manifestation of NFL behavior is the strange metallic phase in cuprate superconductors~\cite{Rev2,Rev4,4,5}, where a $T$-linear scattering rate $1/\tau \sim T$ is observed at quite a wide temperature regime~\cite{np}, while the Landau Fermi liquid theory conventionally predicts that  $1/\tau \sim T^2$ in most metals at low temperatures. As an experimental mystery, the  NFL features like transport properties distinct from the normal Fermi liquid have attracted much attention in the past decades~\cite{Rev2,Rev3,Rev4}. 
Theoretically, the notion of Planckian dissipation, namely the universal Planckian limit on the scattering rate, has been proposed~\cite{Zaanen04,Rev3,Zaanen19}.
Recent numerical studies suggested that this $T$-linear scattering rate can arise in the two-dimensional Hubbard model~\cite{nfl_pnas,triangular}. Specifically, the linear $T$ dependence of electronic scattering rate was discovered to occur in a limited range of doping levels for the square lattice model~\cite{nfl_pnas}. Interestingly, for triangular lattice, the most recent study uncovered two distinct doping regime with different origins of $T$-linear behavior~\cite{triangular}.

One significant issue in the study of unconventional superconductors (SC) is the proper minimal model that 
captures the essential low-energy physics. 
In spite of the success of the one-band Hubbard model and its variants in understanding unconventional SC, their common intrinsic assumption is that the parent compounds, which are
sometimes charge transfer insulators, e.g. cuprate SC\cite{GSA1985}, can instead be modeled as effective Mott insulators.
For this single band Hubbard model, although there is consensus on the absence of SC at hole doping $\delta = 1/8$, it is still highly debated on the existence of SC at other dopings~\cite{hubsc1,hubsc2,hubsc3}. 

Alternatively, the three-orbital Emery model~\cite{emery} provides a more realistic representation of the copper oxide planes as it explicitly incorporates the Cu $d_{x^2-y^{2}}$ and the
two ligand O-2p$_{\sigma}$ orbitals in a unit cell. 
Owing to its more degrees of freedom, the Emery model has been proposed as a way to enhance SC~\cite{Emerysc1,Emerysc2}.
Given that it provides a natural extension so that offers a more accurate depiction of cuprate SC compared to the single-band Hubbard model, it is imperative to ascertain whether it also hosts the linear-in-$T$ behavior of the scattering rate at low temperatures.
Note that due to the complexity originating from the multi-orbital nature, it remains a challenge of achieving consensus on many aspects of its physics~\cite{zhc20,3b_dqmc,3b_num}. 
Furthermore, we believe that the three-orbital Emery model would be particularly important in light of the most recent experimental demonstration of the cuprate-like electronic structure of infinite-layer nickelates~\cite{ding2024cupratelike,sun2024electronic} implying that in some sense the two different families of unconventional SC can be reasonably investigated in a common framework.

To this goal, we have explored the NFL behavior of the electronic scattering rate of the two-dimensional Emery model in different doping levels.  
The additional O degree of freedom introduces one important tuning parameter, namely the site energy $\epsilon_p$ of O-$2p$ orbital compared to the $3d$ orbital ($\epsilon_d=0$ is fixed). Considering that the recent discovered nickelate SC~\cite{2019Nature,Pr,La,LaSC,Nd6Ni5O12} has been proved to have larger charge transfer energy $\Delta \equiv \epsilon_p-\epsilon_d$ than cuprates~\cite{Mi2020,Aritareview,Botana_review}, we focused on two characteristic site energies $\epsilon_p$ of O-$2p$ orbital relevant to cuprates and nickelates respectively to uncover its significant impact on the NFL behavior.
Specifically, our simulations revealed that, for $\epsilon_p$ relevant to cuprates, the electronic scattering rate shows linear-$T$ dependence in a wide range of densities but the interception extrapolated to $T=0$ are always tiny or negative. Intriguingly, the increment of $\epsilon_p$ leads to distinct behavior of the scattering rate, where there exist two consecutive temperature intervals for different slopes of linear-$T$ behavior, which is reminiscent of the recent findings on single band Hubbard model on triangular lattice~\cite{triangular}.
Our investigation would not only deepen our understanding of the fundamental NFL features within Emery model governing the materials like cuprates but also unlock new connections between these two superconducting materials in a common theoretical framework.

\section{Model and Method}

The three-orbital Emery model~\cite{emery,3b_dqmc,3b_num} is defined as 
$H=K_0 + K_{pd} +K_{pp}+V_{dd}+V_{pp}$ with
\begin{align}
    K_0 &= (\epsilon_d - \mu)\sum_{i\sigma} n^d_{i\sigma}+(\epsilon_p-\mu)\sum_{i\alpha\sigma} n_{i\alpha\sigma}^p
     \nonumber\\
    K_{pd} &= \sum_{\langle ij \rangle \alpha  \sigma} t_{pd}^{i,j,\alpha}(d_{i,\sigma}^{\dagger}p_{j,\alpha,\sigma}+p^{\dagger}_{j,\alpha,\sigma}d_{i,\sigma})
    \nonumber\\
    K_{pp} &= \sum_{\langle jj' \rangle \alpha \alpha' \sigma} t_{pp}^{j,j',\alpha,\alpha'}(p^{\dagger}_{j,\alpha,\sigma} p_{j',\alpha',\sigma}+p^{\dagger}_{j',\alpha',\sigma} p_{j,\alpha,\sigma})
    \nonumber\\
    V_{dd} &= U_{dd}\sum_{i} n^d_{i,\uparrow}n^d_{i,\downarrow}
    \nonumber\\
    V_{pp} &= U_{pp}\sum_{j,\alpha}n^p_{j,\alpha,\uparrow}n^p_{j,\alpha,\downarrow}
\end{align}
where $d_{i,\sigma}^{\dagger}$ ($d_{i,\sigma})$ creates (annihilates) a hole with spin $\sigma$ (=$\uparrow$,$\downarrow$) in $d_{x^2-y^2}$ orbital at site i; while $p^{\dagger}_{j,\alpha,\sigma}$ ($p_{j,\alpha,\sigma}$) creates (annihilates) a hole with spin $\sigma$ (=$\uparrow$,$\downarrow$) in the $p_{\alpha}$ ($\alpha$ = x, y) orbital. $n^d_{i\sigma}$ = $d_{i\sigma}^{\dagger}$ $d_{i\sigma}^{\phantom{\dagger}}$ are the number operators; $\langle . \rangle$ means a sum over nearest-neighbor orbitals. $U_{dd}$ and $U_{pp}$ are the strengths of the $d$ and $p$ on-site interactions, respectively. The chemical potential $\mu$ controls the total hole density $\rho$, where $\epsilon_d$ and $\epsilon_p$ are the site energies of the $d$ and $p$ orbitals respectively. $t_{pd}^{ij\alpha}= t_{pd}(-1)^{\eta_{ij}}$ and $t_{pp}^{jj'\alpha\alpha'}= t_{pp}(-1)^{\beta_{jj'}}$ are the nearest-neighbor $d$-$p$ and $p$-$p$ hopping integrals. In the hole language, ${\eta_{ij}}$ and ${\beta_{jj'}}$ take values $\pm$1 following the conventions. In hole language, the phase convention is $\eta_{ij} = 1$ for $j = i + \frac{1}{2}x, \alpha = x$ or $j = i - \frac{1}{2}y, \alpha = y$ and $\eta_{ij} = 0$ for $j = i - \frac{1}{2}x, \alpha = x$ or $j = i + \frac{1}{2}y, \alpha = y$. In addition, $\beta_{jj'} = 1$ for $j' = j - \frac{1}{2}x - \frac{1}{2}y$ or $j' = j + \frac{1}{2}x + \frac{1}{2}y$ and $\beta_{jj'} = 0$ for for $j' = j - \frac{1}{2}x + \frac{1}{2}y$ or $j' = j + \frac{1}{2}x - \frac{1}{2}y$, $\alpha = x$ and $\alpha^{'}=y$ or $\alpha = y$ and $\alpha^{'} = x$, respectively.
Other conventions are also applicable due to the gauge invariance~\cite{3b_dqmc}. Unless otherwise stated, we use the parameters listed below (in units of eV): $U_{dd}$ = 7.5, $U_{pp}$ = 0, $t_{pd}$ = 1.13, $t_{pp}$ = 0.49, $\epsilon_d$ = 0. Note that we do not adopt $U_{dd}=8.5$ in the literature to alleviate the sign problem in our simulations with large DCA cluster $N_c=16$, which should not qualitatively modify our results presented here.

In this work, we have endeavored to solve the two-dimensional three-orbital Emery model at low temperatures using the dynamical cluster approximation (DCA)~\cite{Hettler98,Maier05,code} with the continuous-time auxilary-field (CT-AUX) quantum Monte Carlo (QMC) cluster solver~\cite{GullCTAUX}.
As an advanced quantum many-body numerical method, DCA evaluates the physical quantities in the thermodynamic limit via mapping the bulk lattice problem onto a finite cluster embedded in a mean-field bath in a self-consistent manner~\cite{Hettler98,Maier05}. More discussions on DCA technique and its insight on the strongly correlated electronic systems can be found in Ref.~\cite{Maier05}.

Our focused physical quantity is the electronic scattering rate $\gamma_{k}\equiv$ -Im$\Sigma(\mathbf{K},\omega = 0)$, which is extracted from a $m$-th order polynomial extrapolation of -Im$\Sigma^{(m)}(\mathbf{K},i\omega_n)$ in lowest Matsubara frequencies. 
Although the $m$ value can affect the extrapolation generically, our results show qualitatively similar trend of $T$ dependence at low temperatures so that normally $m=2$ is chosen~\cite{nfl_pnas,triangular}. Note that, as an approximation avoiding the ambiguous and challenging analytical continuation procedure, the accuracy of this extrapolation for zero frequency improves at low temperatures where the Matsubara frequencies are closer. 

Most of our calculations were conducted with $N_c=$16 sites DCA cluster for fine enough but still computationally manageable momentum space resolution including nodal $\mathbf{K}= (\pi/2,\pi/2)$ and antinodal $\mathbf{K} = (\pi,0)$ directions.
Despite of its accuracy, the relatively large $N_c=16$ does not allow accessing low enough temperatures due to QMC sign problem so that some simulations using $N_c=4$ provide further insights on the lowest temperature features in spite of the lack of the self energy at the nodal direction.
Fortunately, different $N_c$ lead to quite similar results for high dopings (large $\rho$) and large $\epsilon_p$. At lower dopings, however, it is not the case anymore so that adopting small $N_c=4$ can lead to deviations from the physical reality and require more careful examination.

Another aspect is the momentum differentiation of the scattering rate, namely the deviation between nodal and antinodal directions, which is conventionally associated with the pseudogap (PG) features~\cite{nfl_pnas}. It is also valuable to explore the local scattering rate $\Gamma$ as the momentum average of $\gamma_k$~\cite{triangular}. We believe that this is worthwhile even in the anisotropic situations to explore the difference between momentum averaged scattering rate and the values for a particular $\mathbf{K}$ direction. 
Throughout this work, we focus on the temperature evolution of $\Gamma$ and $\gamma_k$ for various hole density $\rho$ (note that $\rho > 1$ corresponds to the hole doping). Regarding the site energy $\epsilon_p$, we do not restrict on the case with $\epsilon_p = 3.24$ eV specific to cuprates but extend it to larger value e.g. $\epsilon_p \sim 6.0$ eV relevant to different compounds e.g. nickelates~\cite{Mi2020,Aritareview,Botana_review} or physical situations.

\section{Results}
\subsection{$\epsilon_p=3.24$ eV for cuprates}

\begin{figure*} 
\psfig{figure=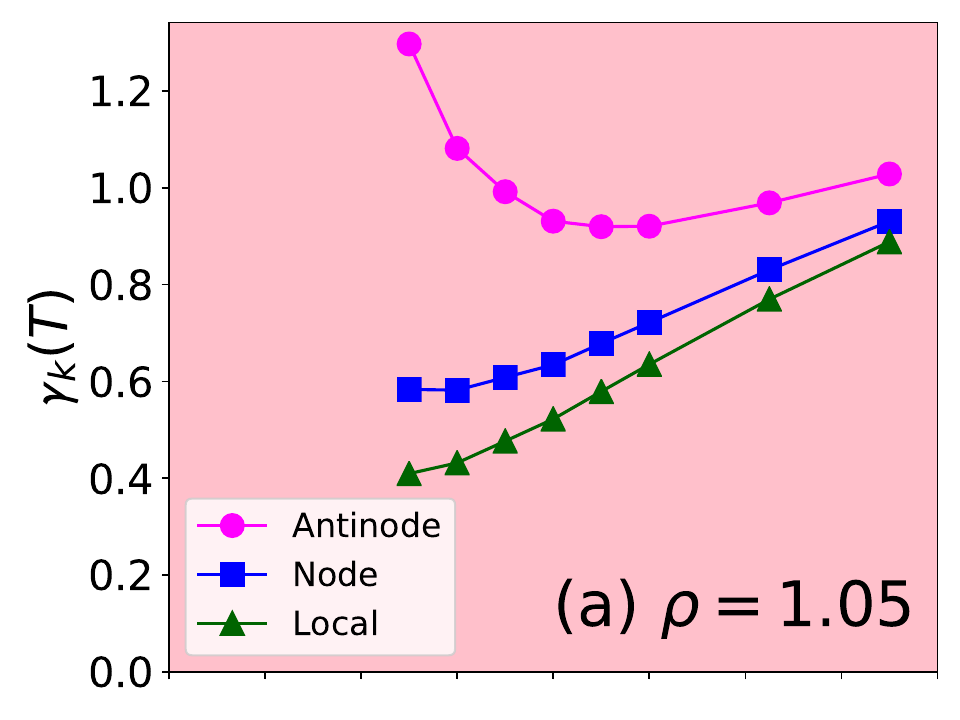, height=2.8cm,width = .23\textwidth, trim={0 0 0 0}}
\psfig{figure=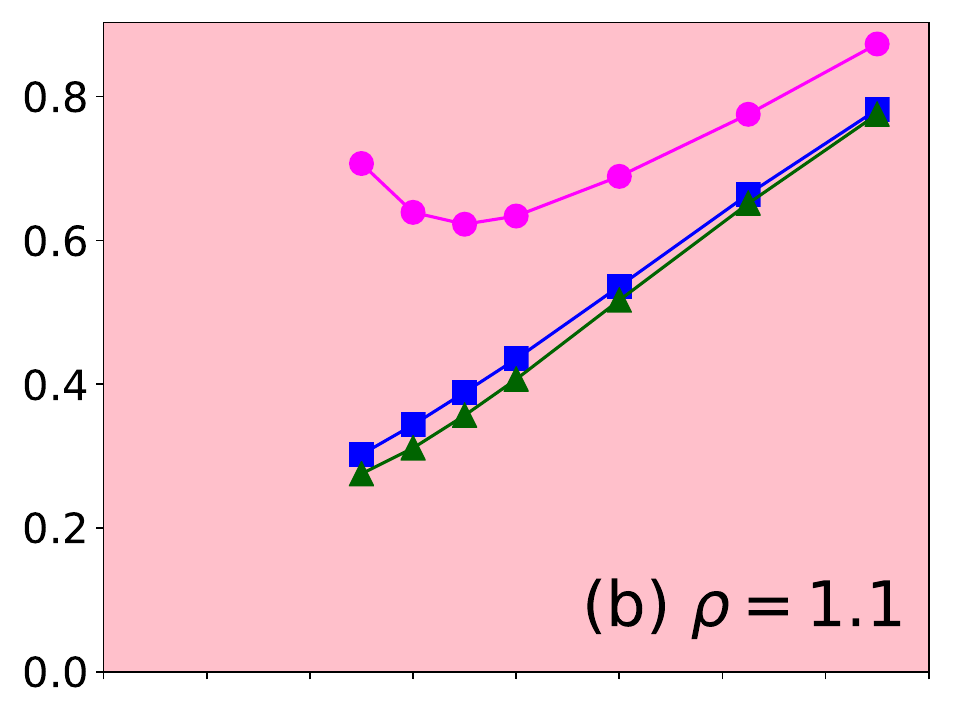, height=2.8cm,width = .23\textwidth, trim={0 0 0 0}}
\psfig{figure=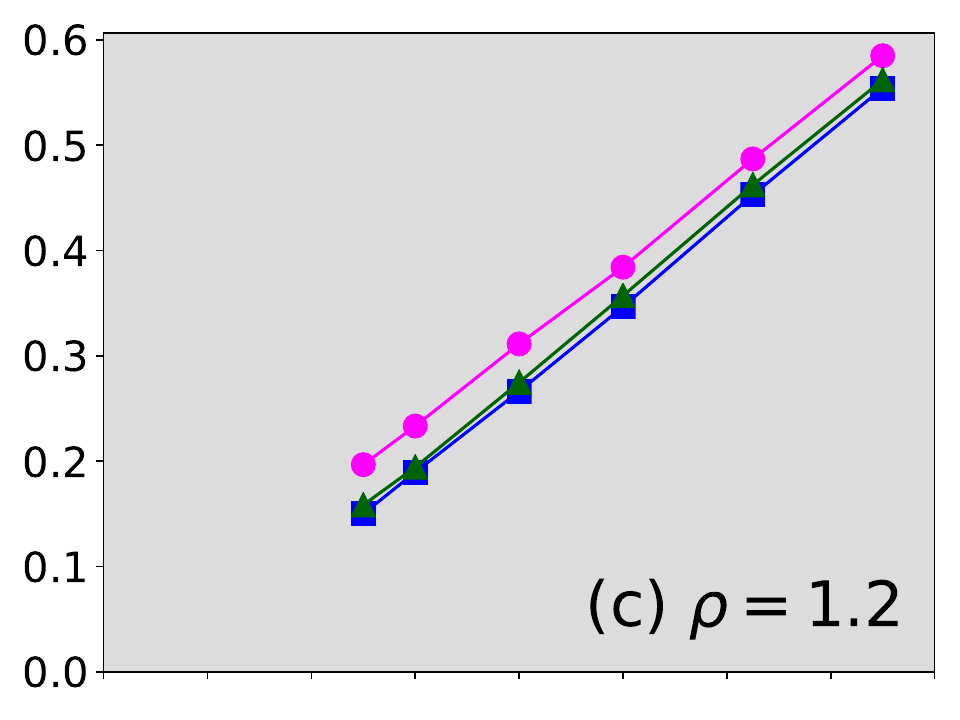, height=2.8cm,width = .23\textwidth, trim={0 0 0 4}}
\psfig{figure=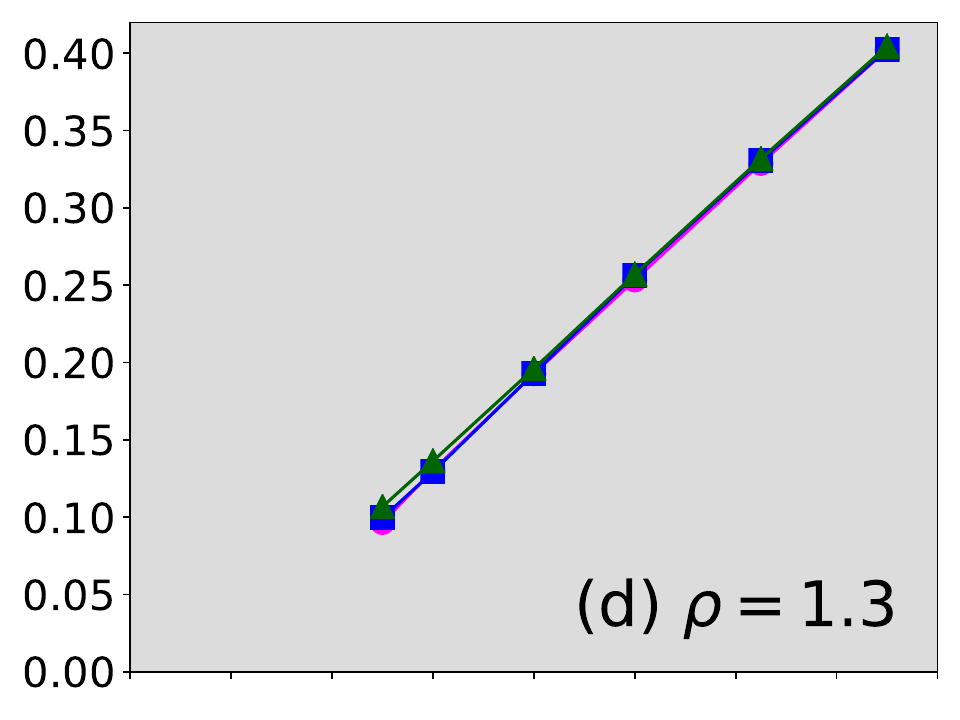, height=2.8cm,width = .23\textwidth, trim={0 0 0 0}}
\psfig{figure=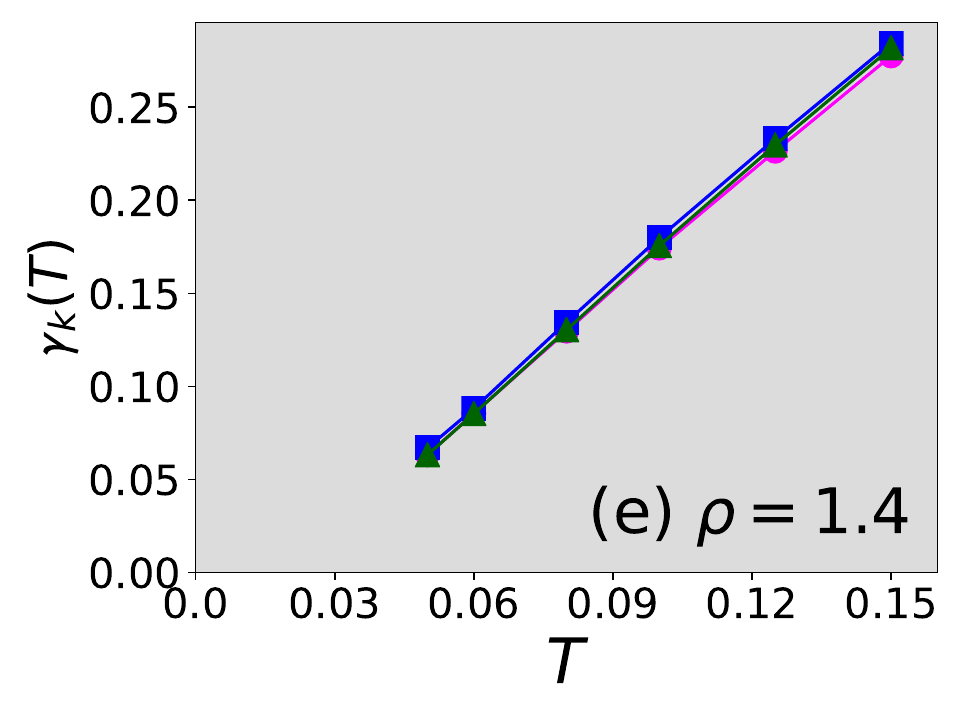, height=3.0cm,width = .23\textwidth, trim={14 0 0 0}}
\psfig{figure=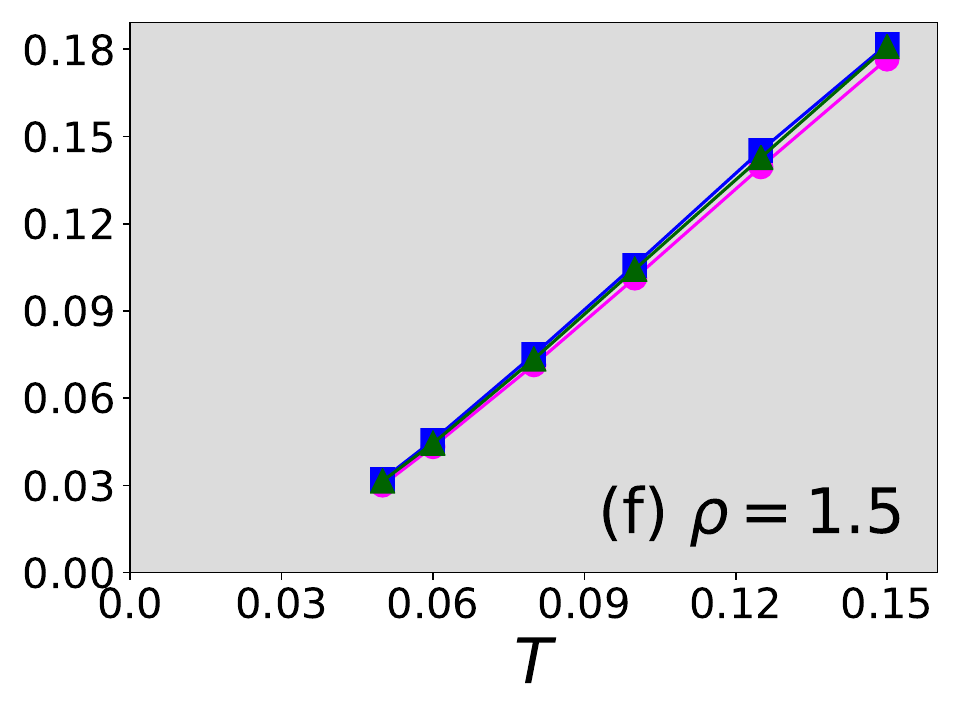, height=3.0cm,width = .23\textwidth, trim={12 0 0 0}}
\psfig{figure=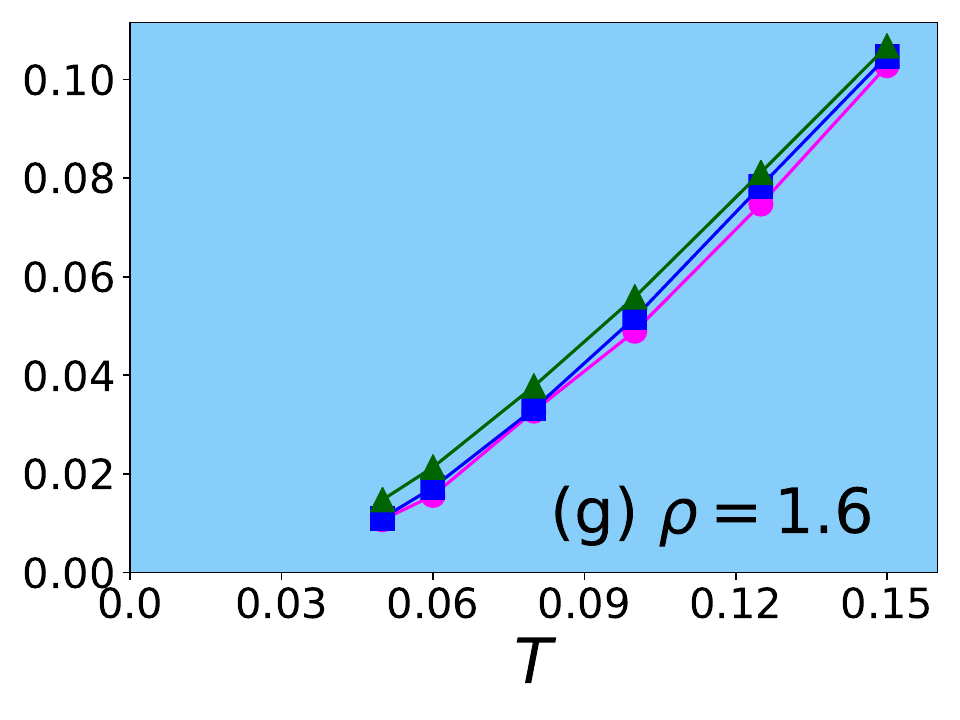, height=3.0cm,width = .23\textwidth, trim={12 0 0 0}}
\psfig{figure=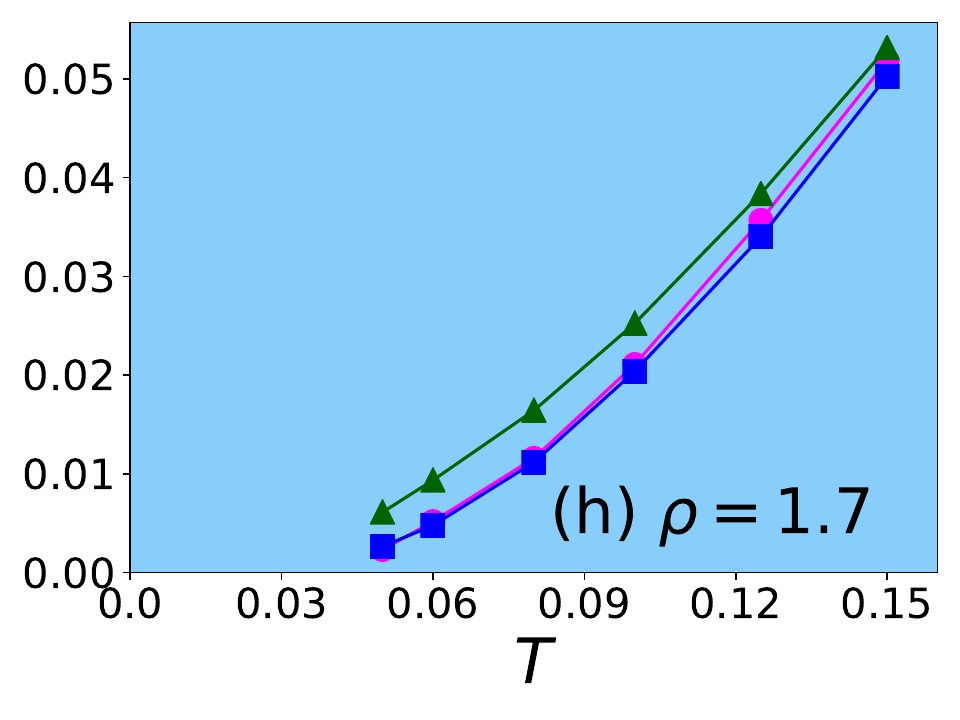, height=3.0cm,width = .23\textwidth, trim={0 0 0 0}}
\psfig{figure=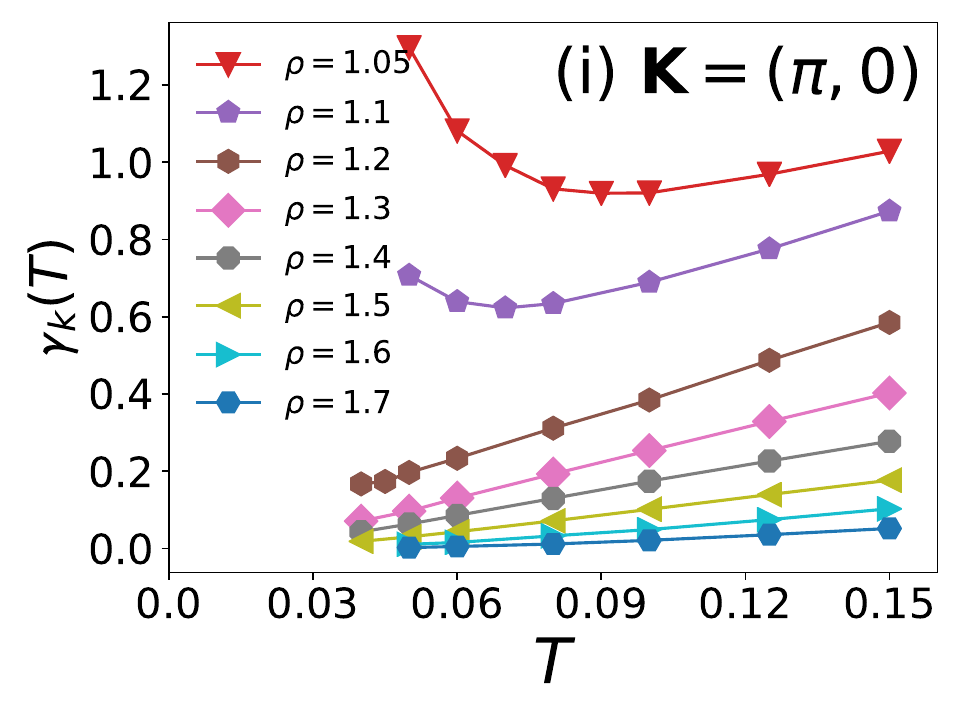, height=3.0cm,width = .23\textwidth, trim={0 0 0 0}}
\psfig{figure=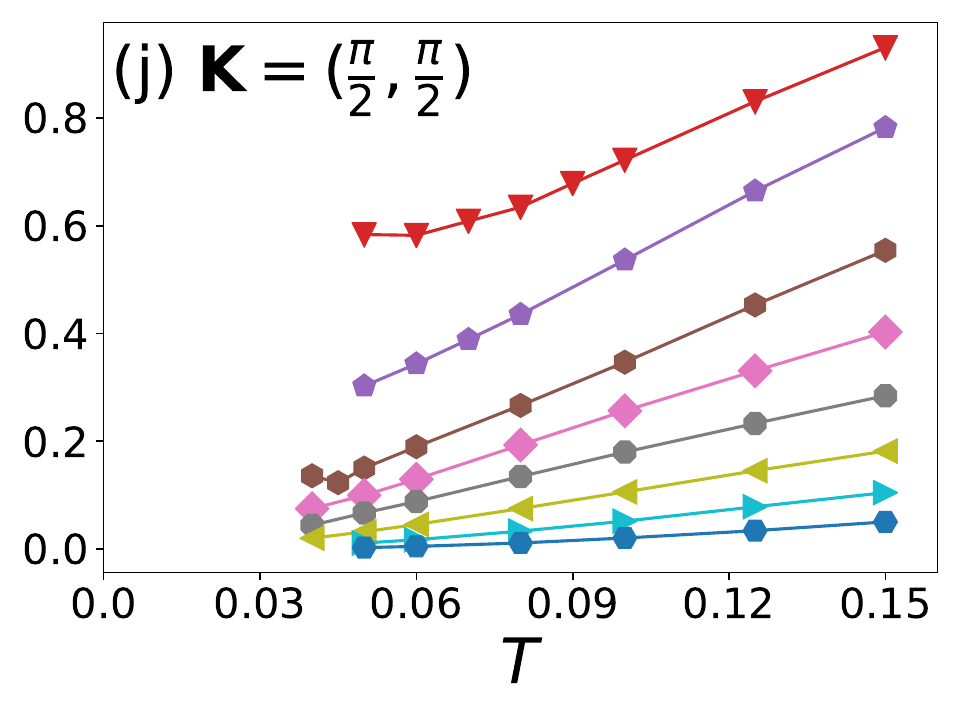, height=3.0cm,width = .23\textwidth, trim={0 0 0 0}}
\psfig{figure=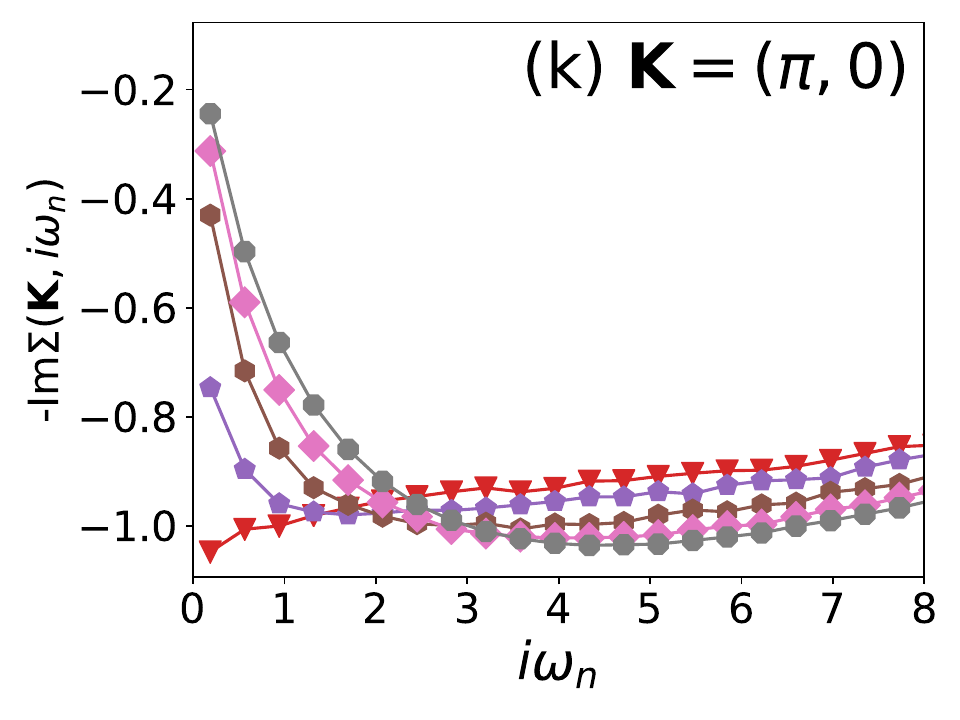, height=3.0cm,width = .23\textwidth, trim={20 0 0 0}}
\psfig{figure=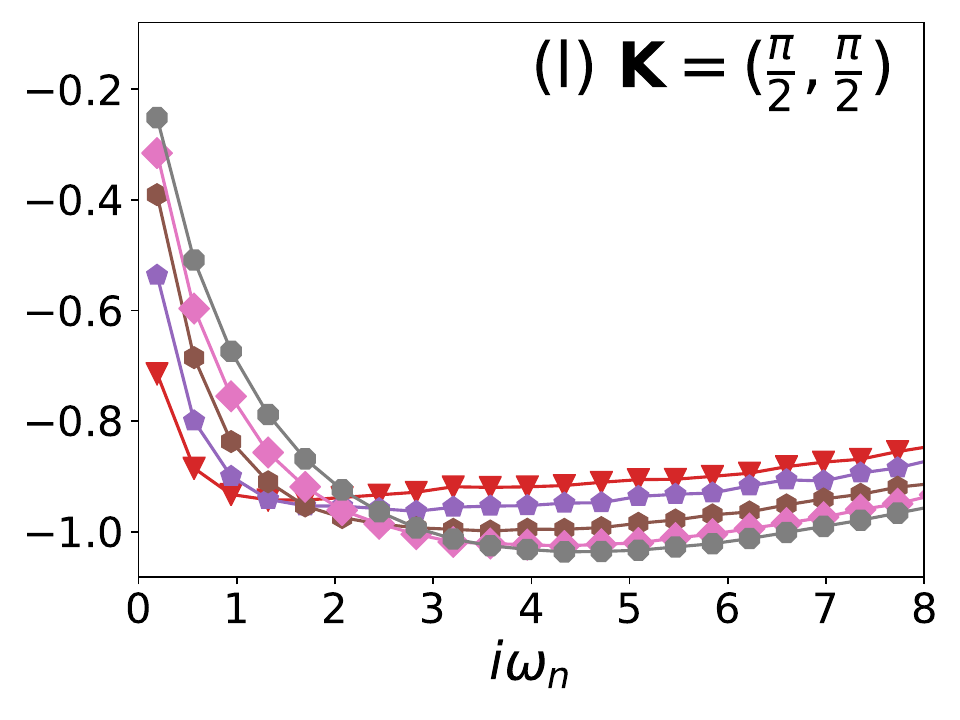, height=3.0cm,width = .23\textwidth, trim={0 0 0 0}}
\caption{(a-h) Temperature dependence of the electronic local scattering rate $\Gamma$ and momentum-resolved $\gamma_k$ at nodal $(\pi/2,\pi/2)$ and antinodal $(\pi,0)$ directions for various density at $\epsilon_p=3.24$ relevant to cuprates;
(i-j) Comparison of $\gamma_k$ with varying densities; (k-l) Frequency dependence of -Im$\Sigma(\mathbf{K},i\omega_n)$ for different densities at $T=0.06$.}
\label{fig1}
\end{figure*}

Figure~\ref{fig1} illustrates the temperature dependence of local scattering rate $\Gamma$ as well as momentum-resolved $\gamma_k$ of $d$ orbital for varying density $\rho$ at $\epsilon_p=3.24$ relevant to cuprates. 
At small $\rho=1.05$ (small hole doping) in panel (a), both the nodal and antinodal $\gamma_k$ show a prominent upturn at low temperatures indicating the insulating behavior due to its closeness to the charge transfer insulator at half-filling $\rho=1.0$. 
This trend changes to the typical pseudogap (PS) feature akin to the single band model~\cite{nfl_pnas}, which is manifested by the observation that only the antinodal $\gamma_k$ shows the upturn while the nodal $\gamma_k$ remains its monotonic evolution until lowest simulated temperature, for instance, as shown in panel (b).

As the doping becomes progressively heavier (when $\rho$ reaches 1.2 or higher), the linear $T$ dependence of $\gamma_k$ at both directions extend to $T\rightarrow0 $. 
Our simulation indicate that the slope of $\gamma_k(T)$ at two directions are almost the same in a wide range of densities as evidenced in panels (c-f). Interestingly, further hole doping at $\rho>1.25$ leads to the nearly isotropic $\gamma_k(T)$, namely independent on the $\mathbf{K}$ direction. This is evidenced by the overlap of the local scattering rate $\Gamma(T)$ with the momentum-resolved $\gamma_k(T)$ for intermediate density range as shown in panels (d-f).  
It is discovered that the linear-in-$T$ scattering rate persists for quite a wide density regime around $\rho \sim 1.2-1.5$, which extends to heavily doped side. Only at even higher density $\rho >1.5$, $\gamma_k(T)$ deviates from the linear evolution at low temperatures. Note that in fact we cannot determine the physical behavior of $\gamma_k(T)$ at even lower temperatures, where the evolution can change in a qualitative manner.
One additional interesting feature lies in the behavior of the local scattering rate $\Gamma$ at high density, where it deviates from the overlapped $\gamma_k$ of nodal and antinodal directions, which arises from the slightly different behavior at other directions e.g. $\mathbf{K}=(0,0), (\pi,\pi)$. Therefore, the truly isotropic scattering rate only applies for the intermediate density regime. 

The caveat here is that physically the scattering rate should remain non-negative. Nonetheless, the interception of the linear fit (extrapolating $\gamma_k$ to zero via $\gamma_k \sim aT+b$) leads to a non-physical $\gamma_k(T=0) < 0$. In fact, our simulated curves all show quite small positive or even negative $\gamma_k(T=0)$. Hence, higher order corrections such as quadratic evolution might develop in $\gamma_k$ at lower $T$ as the indication of the onset of Fermi liquid physics or more complicated NFL features. Unfortunately, our simulations are limited by the severe sign problem for large enough $N_c$ so that the physics at $T \rightarrow 0 $ is not accessible at least for DCA simulations. 
As discussed previously in the literature~\cite{nfl_pnas,triangular}, it is not appropriate to directly designate the observed behavior as the strange metal in these cases.

More detailed comparison between various situations are displayed in the bottom row of Fig.~\ref{fig1}. The left two panels provide evidence that the slope of linear-$T$ scattering rate $\gamma_k(T)$ decreases monotonically with increasing density. In fact, the scale of $\gamma_k(T)$ is already quite small at $\rho=1.5$, indicating the strong metallic nature owing to the heavily doped charge carriers, in spite of its linear-$T$ behavior. 
For completeness, the bottom right two panels give the frequency dependence of the self-energy, where the scattering rate $\gamma_k(T)$ is extracted. The transition from the low density momentum differentiation to high density isotropy is obvious.

\begin{figure*} 
\psfig{figure=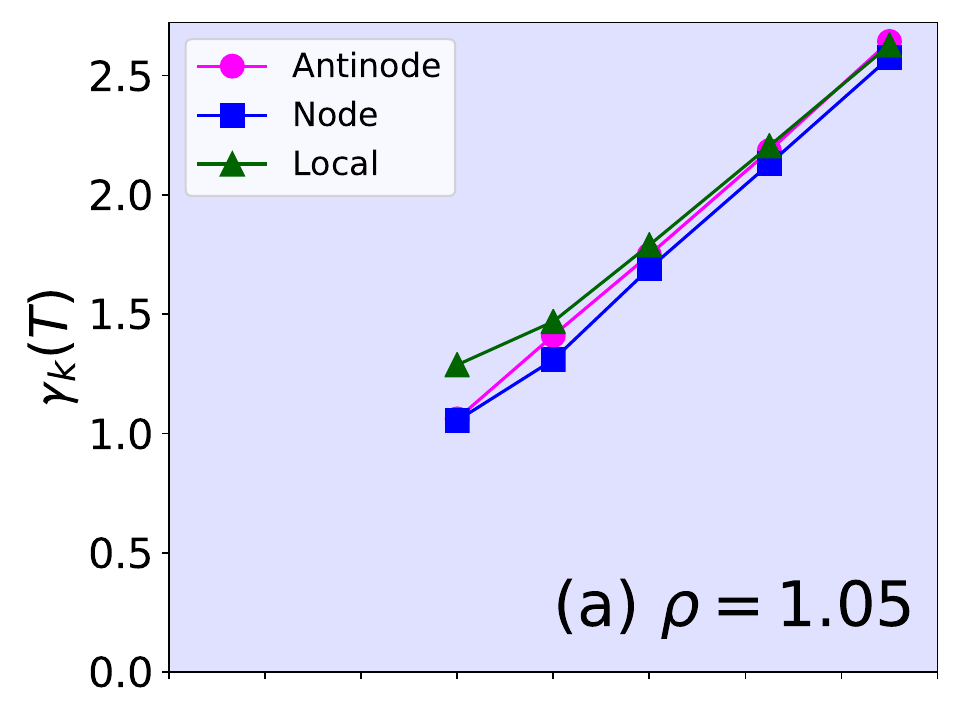, height=2.8cm,width = .23\textwidth, trim={0 0 0 0}}
\psfig{figure=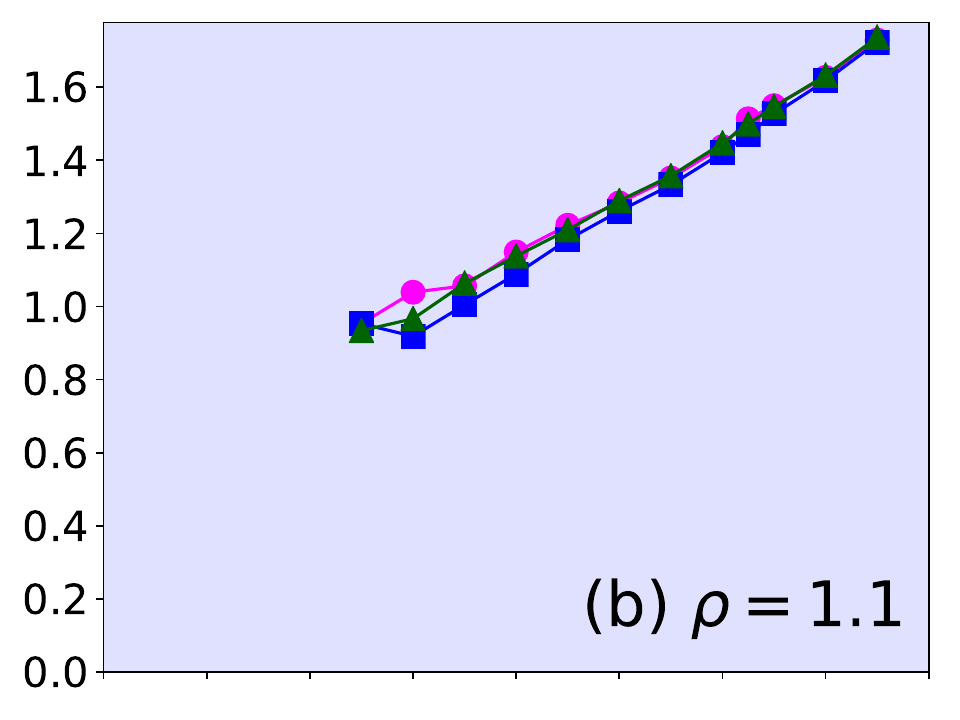, height=2.8cm,width = .23\textwidth, trim={0 0 0 0}}
\psfig{figure=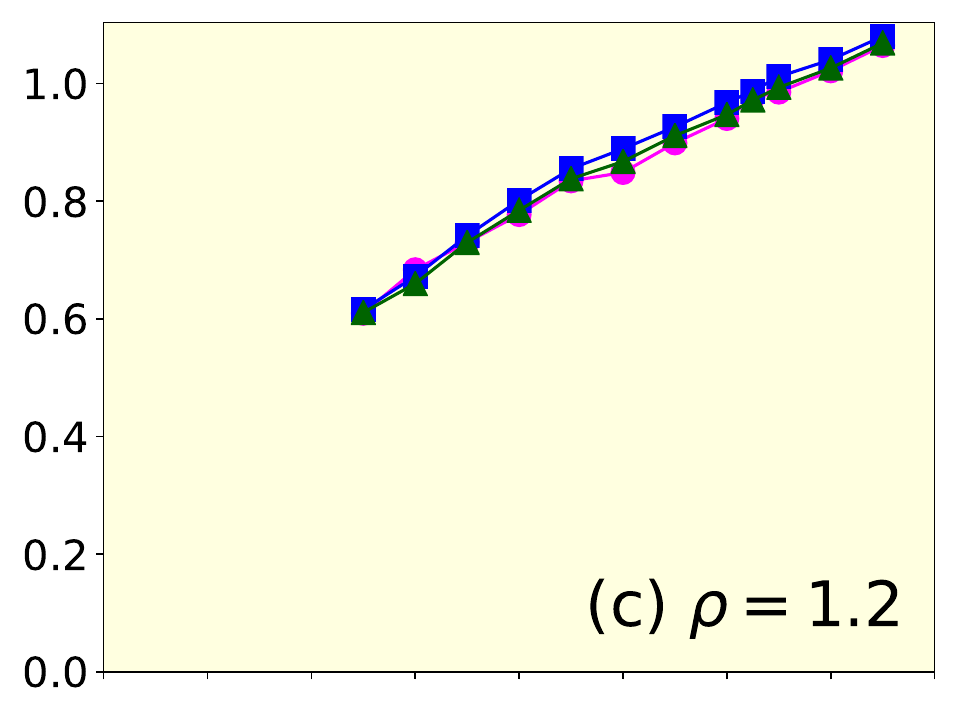, height=2.8cm,width = .23\textwidth, trim={0 0 0 0}}
\psfig{figure=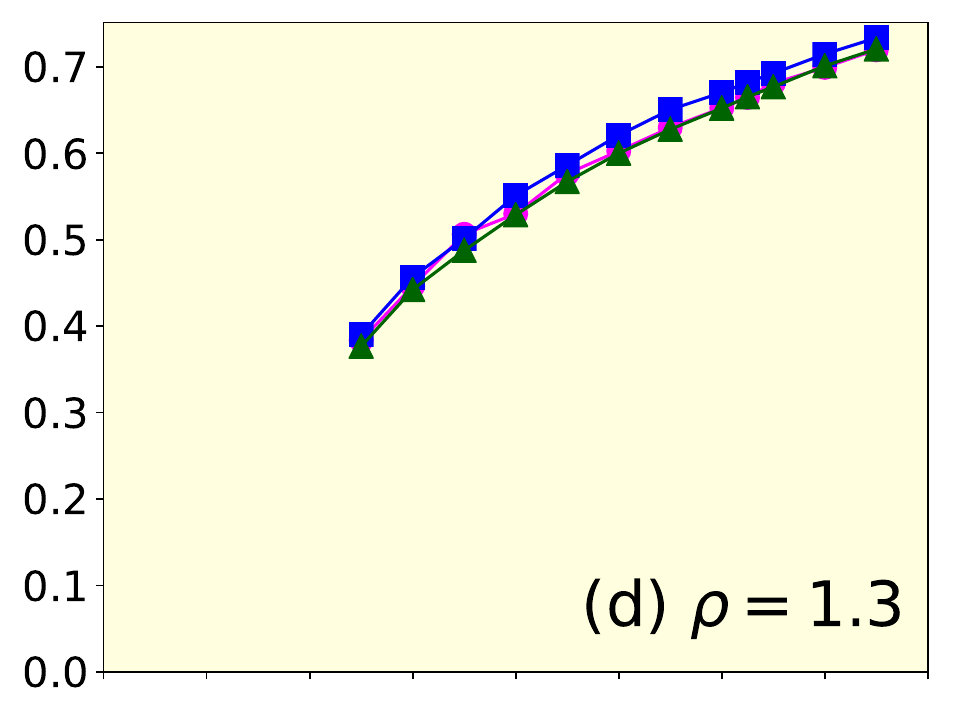, height=2.8cm,width = .23\textwidth, trim={0 0 0 0}}
\psfig{figure=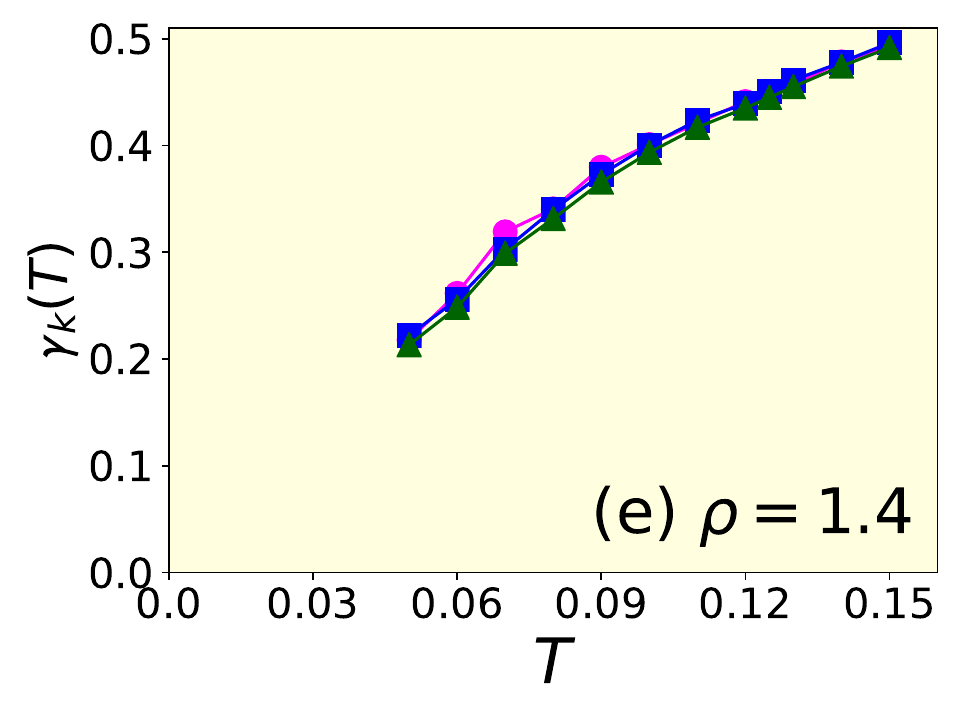, height=3.0cm,width = .23\textwidth, trim={0 0 0 3}}
\psfig{figure=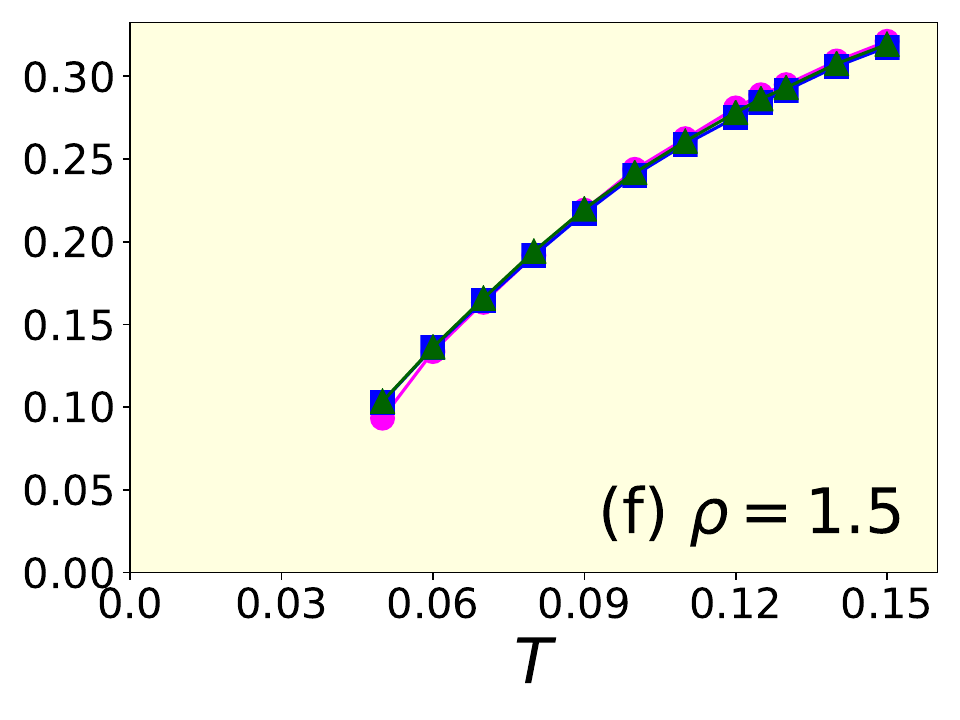, height=3.0cm,width = .23\textwidth, trim={12 0 0 0}}
\psfig{figure=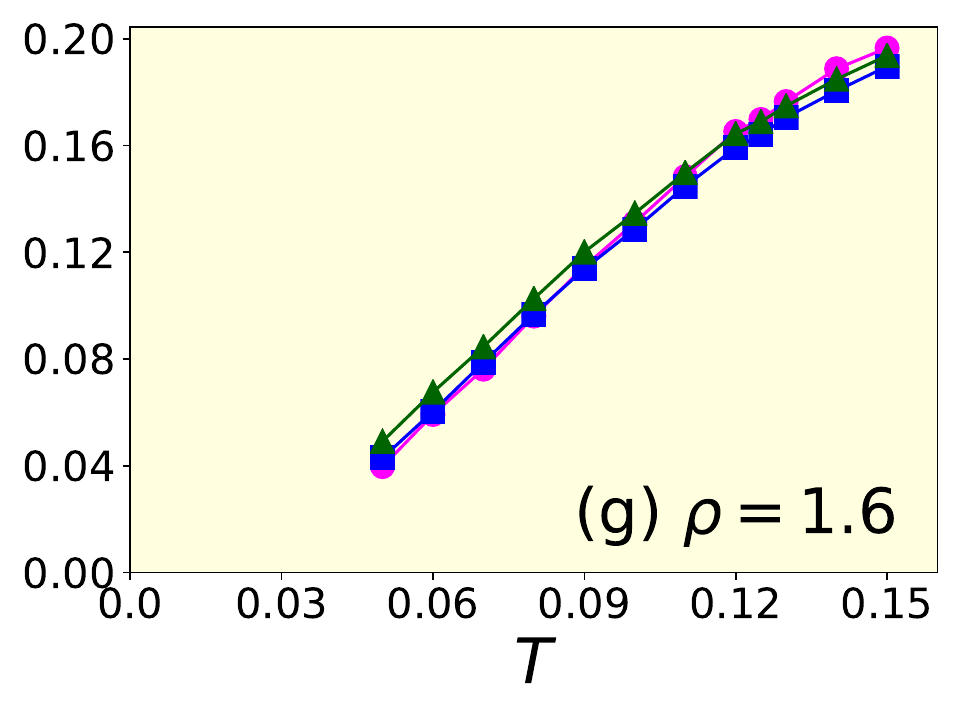, height=3.0cm,width = .23\textwidth, trim={12 0 0 0}}
\psfig{figure=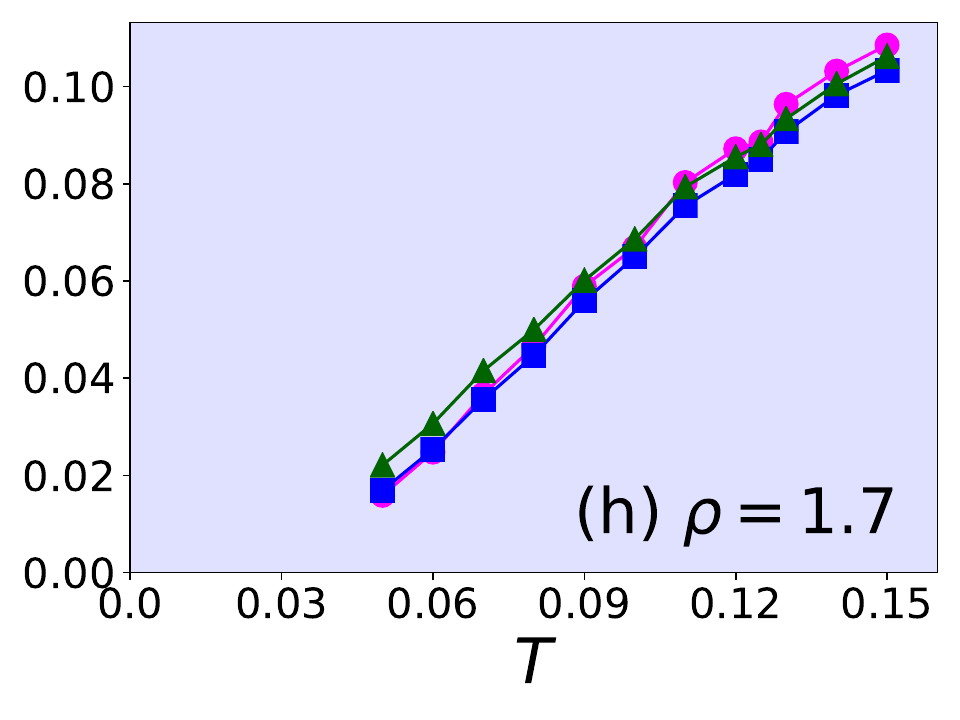, height=3.0cm,width = .23\textwidth, trim={0 0 0 0}}
\psfig{figure=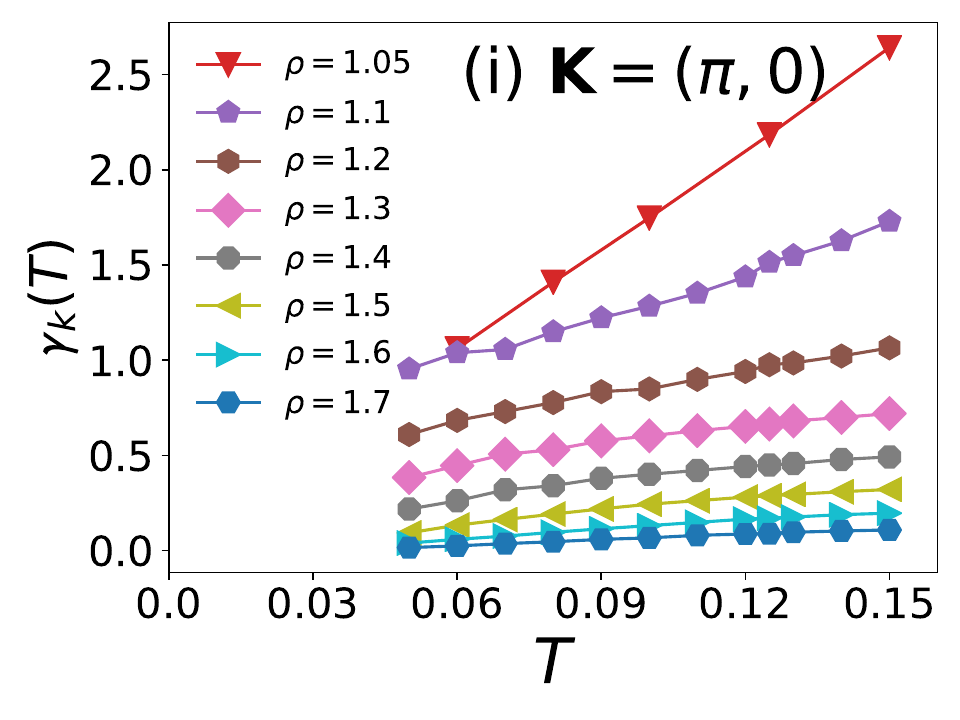, height=3.0cm,width = .23\textwidth, trim={0 0 0 0}}
\psfig{figure=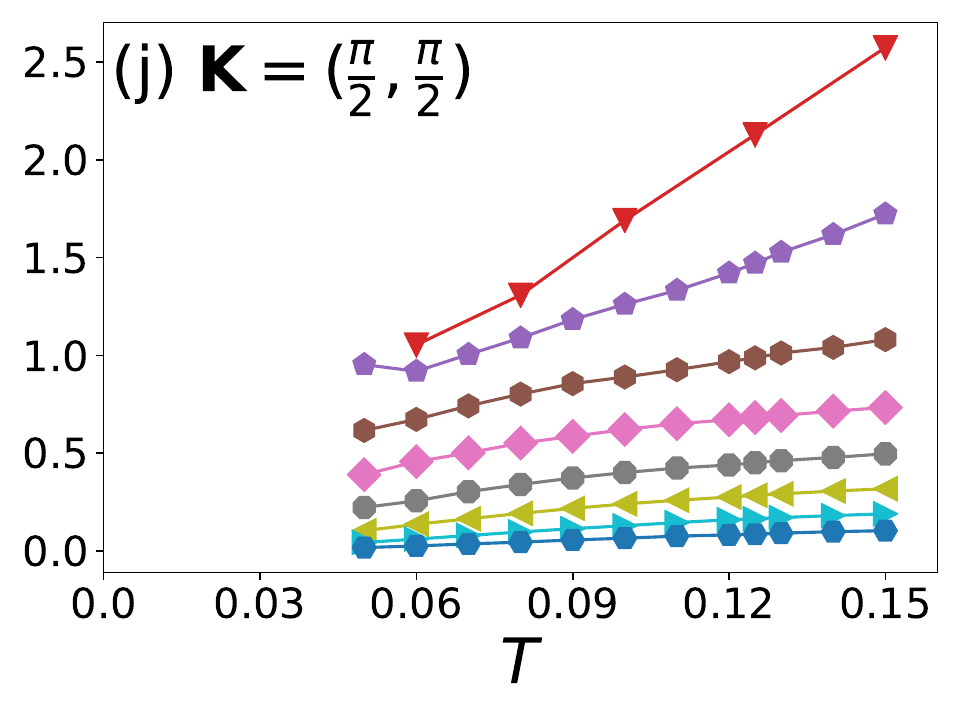, height=3.0cm,width = .23\textwidth, trim={0 0 0 0}}
\psfig{figure=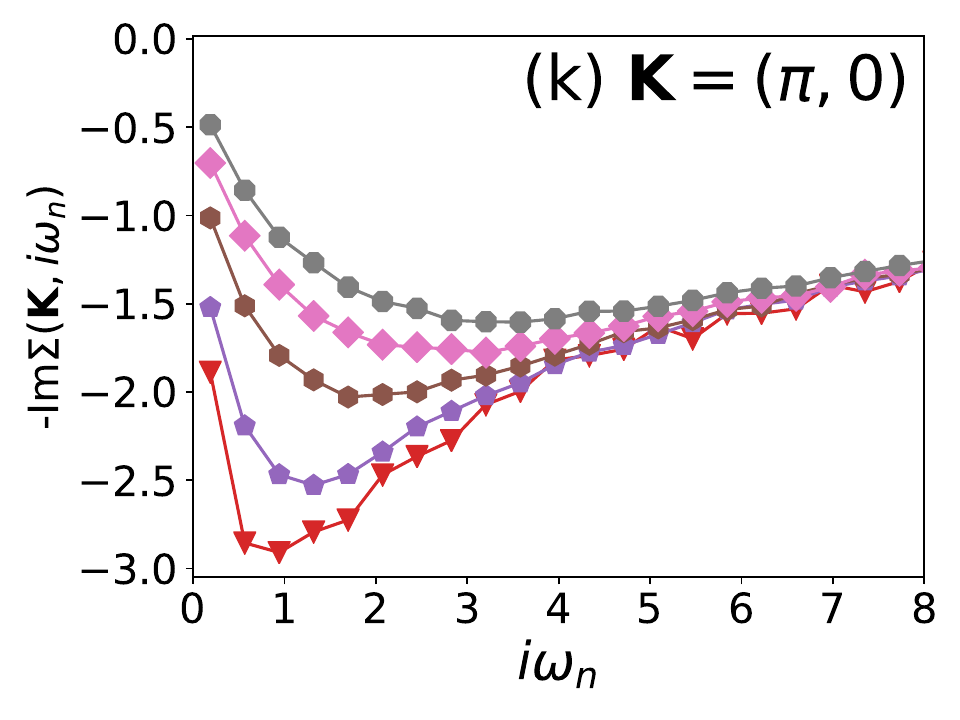, height=3.0cm,width = .23\textwidth, trim={20 0 0 0}}
\psfig{figure=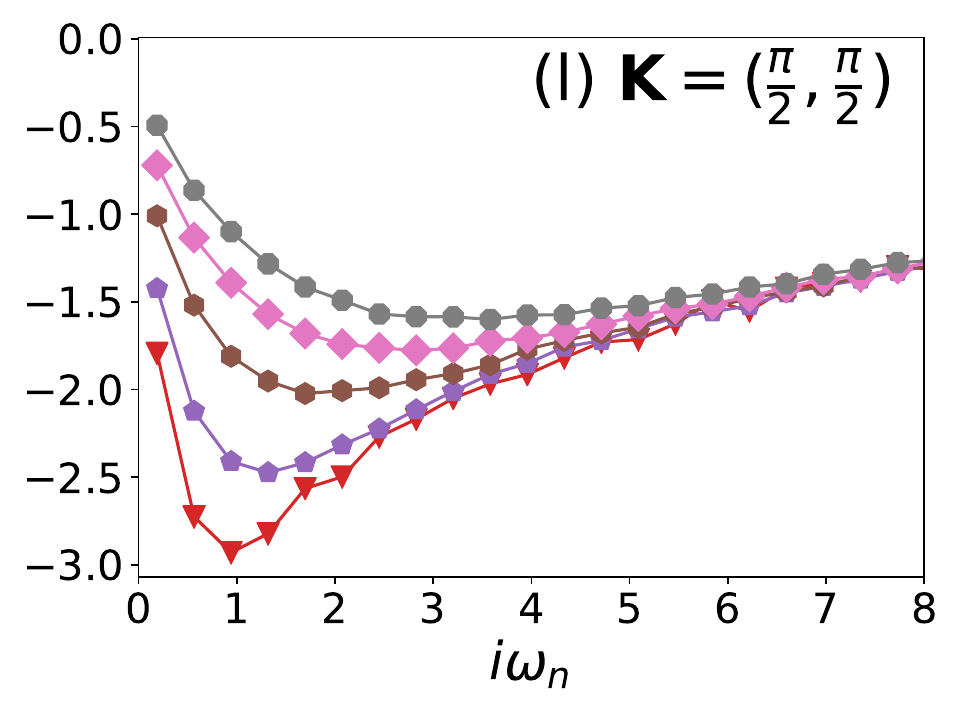, height=3.0cm,width = .23\textwidth, trim={0 0 0 0}}

\caption{Temperature dependence of the electronic local scattering rate $\Gamma$ and momentum-resolved $\gamma_k$ at nodal $(\pi/2,\pi/2)$ and antinodal $(\pi,0)$ akin to Fig.~\ref{fig1} while at $\epsilon_p=6.0$ presumably relevant to nickelates.}
\label{fig2}
\end{figure*}

\subsection{$\epsilon_p=6.0$ eV for nickelates}

As mentioned before, our study does not restrict on the parameter sets relevant to cuprates. Figure~\ref{fig2} demonstrates the same scattering rates and frequency dependent self-energy with only different $\epsilon_p=6.0$ eV closely related to the recently discovered nickelate SC~\cite{Mi2020,Botana_review}.

\begin{figure} 
\psfig{figure=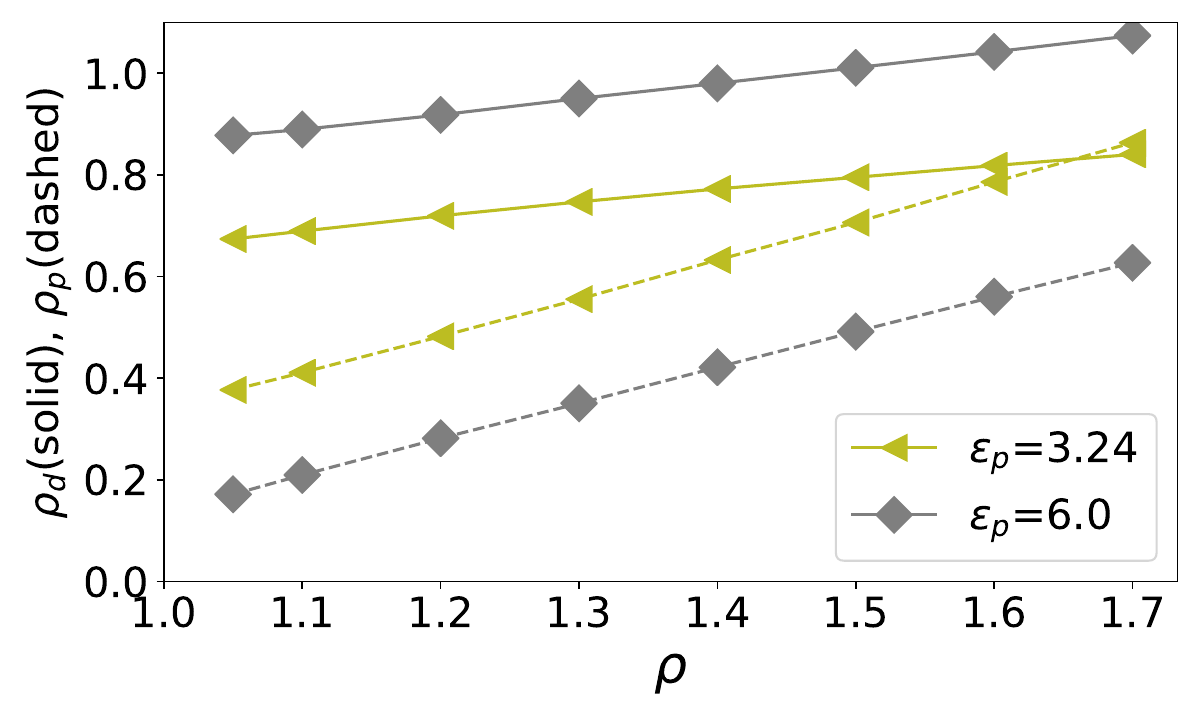, height=5cm,width = .45\textwidth, trim={0 0 0 0}}
\caption{Evolution of charge distribution with total density $\rho$. Larger $\epsilon_p=6.0$ promotes more hole occupancy on $d$ orbital compared to $\epsilon_p=3.24$ while suppresses $\rho_p$.}
\label{ndnp}
\end{figure}

Firstly, the distinction from the $\epsilon_p=3.24$ situation is that, at the same density, $\gamma_k$ is globally larger than the values for $\epsilon_p=3.24$, indicating stronger interaction effects. 
Note that our model does not include the explicit $U_{pp}$ (to avoid the severe sign problem but its role needs further exploration) so that the sole player governing the electronic interaction seems originating from $U_{dd}$ while it normally leads to momentum differentiation, whose absence in our simulations prompts additional reasons for the observed larger scattering rates. 
Physically, the large $\epsilon_p$ discourages the charge carriers locating onto the $p$ orbitals so that the effectively more carrier density on $d$ orbital induces stronger interaction effects from $U_{dd}$. The charge redistribution with different $\epsilon_p$ can be clearly seen in Fig.~\ref{ndnp}, where the larger $\epsilon_p=6.0$ promotes more hole occupancy on $d$ orbital compared to $\epsilon_p=3.24$ while suppresses $\rho_p$. Note also the comparison of the increasing rate of $\rho_d(\rho)$ versus $\rho_p(\rho)$ for two different $\epsilon_p$.

Secondly, $\gamma_k$ is almost isotropic with respect to $\mathbf{K}$ regardless of the density except for very low hole doping $\rho=1.05$ in panel (a). This is further evidenced by the overlap between the local scattering rate $\Gamma$ and the momentum-resolved $\gamma_k$ in most cases.
Owing to this observation, it is plausible to examine the fate of $\gamma_k$ at lower temperature by adopting smaller $N_c=4$ DCA cluster, which will be discussed later.

The most prominent feature for $\epsilon_p=6.0$ is that the linear-$T$ dependence of $\gamma_k$ occurs only for a limited temperature interval at small density like $\rho=1.05$ in panel (a) or large density $\rho=1.7$ in panel (h) in spite of the extrapolated negative interception at $T=0$. 
In contrast, for intermediate densities, $\gamma_k$ exhibits a downturn as the temperature decreases as shown in panels (c-g). This phenomena is similar to the recent findings for single band Hubbard model in triangular lattice that is claimed to originate from two types of mechanism~\cite{triangular}.
Note that our observed downturn starts from a slightly higher temperature scale $T\sim 0.1$.

Akin to Fig~\ref{fig1}, the two left bottom panels provide summarized comparison between  $\gamma_k$ at the antinodal and nodal directions. The trend of decreasing scattering rate with hole doping is apparently similar to the situations for $\epsilon_p=3.24$. The two right bottom panels vividly show the isotropy of the self-energy in systems of large $\epsilon_p=6.0$.

\subsection{Downturn of $\gamma_k$ at large $\epsilon_p$}

\begin{figure}
\psfig{figure=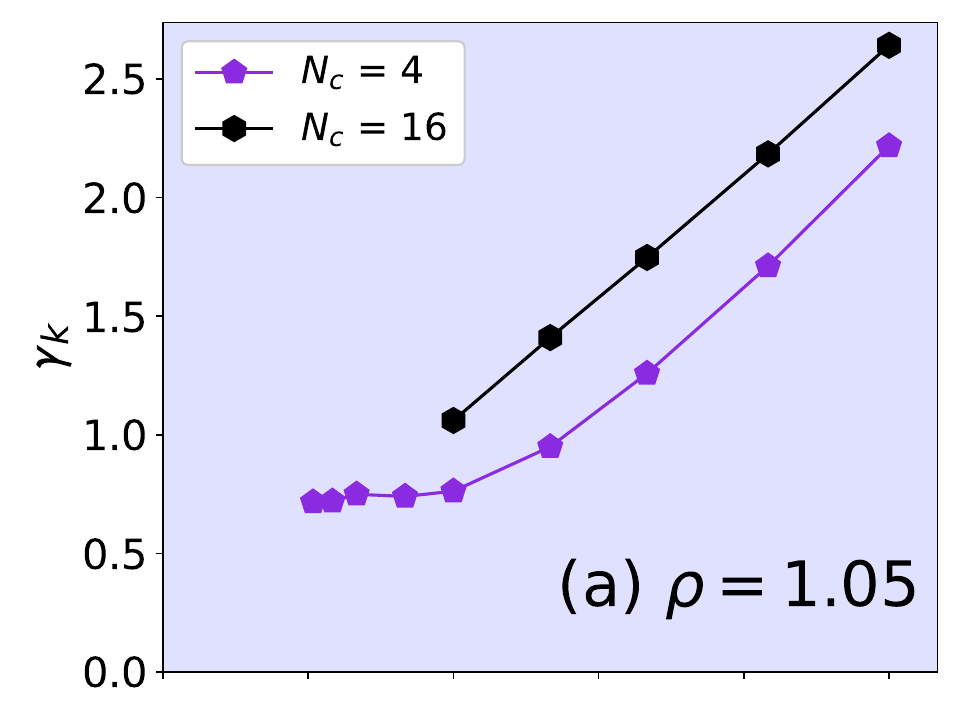, height=2.8cm,width = .23\textwidth, trim={0 0 0 0}}
\psfig{figure=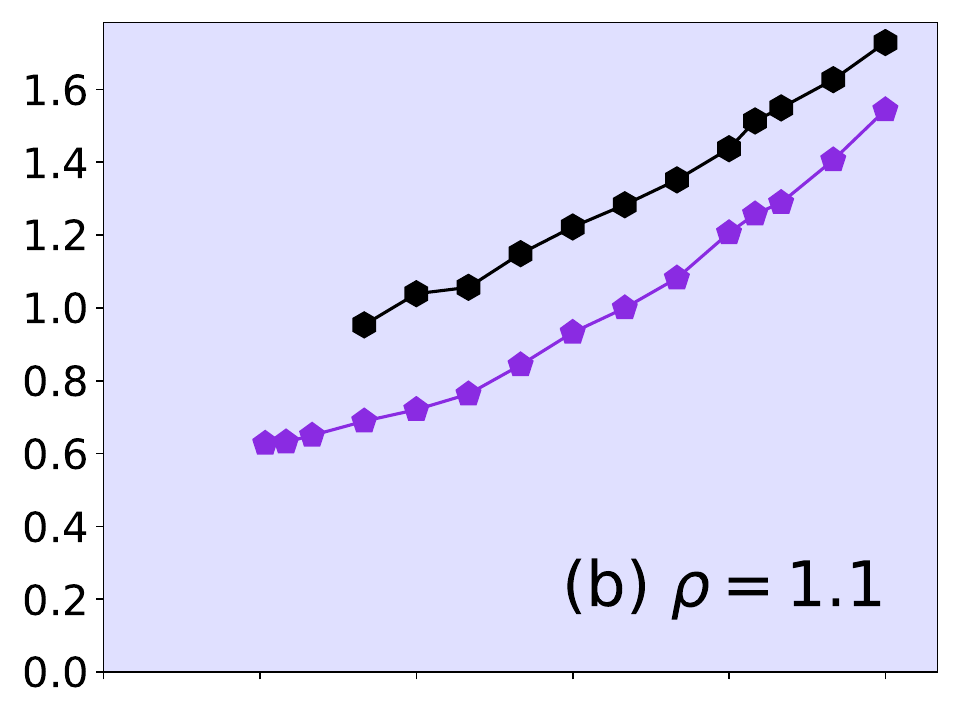, height=2.8cm,width = .23\textwidth, trim={0 0 0 0}}
\psfig{figure=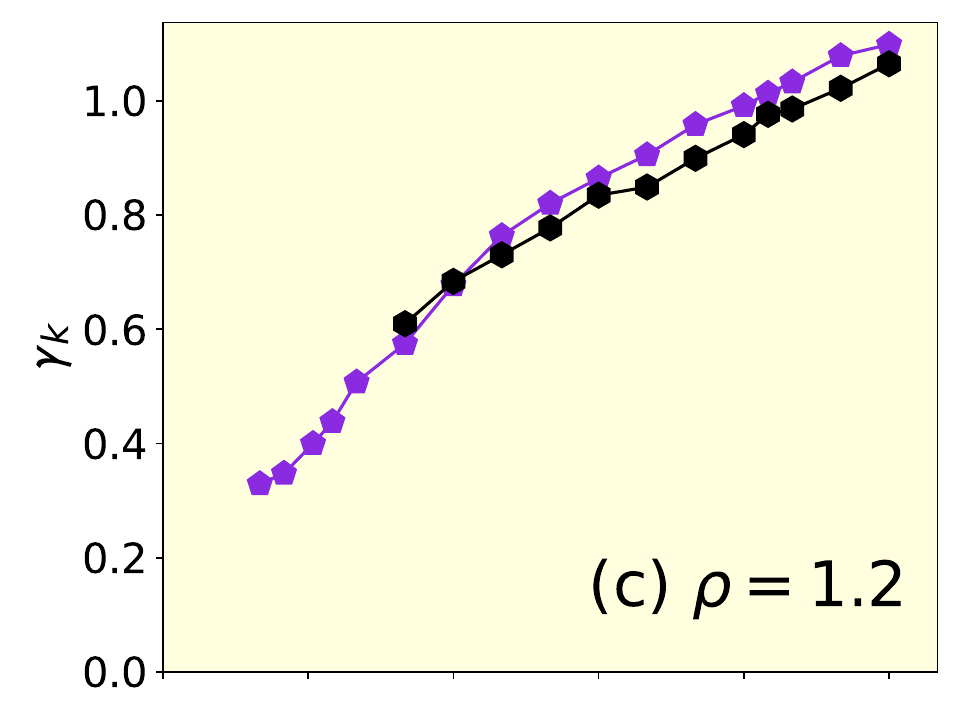, height=2.8cm,width = .23\textwidth, trim={0 0 0 0}}
\psfig{figure=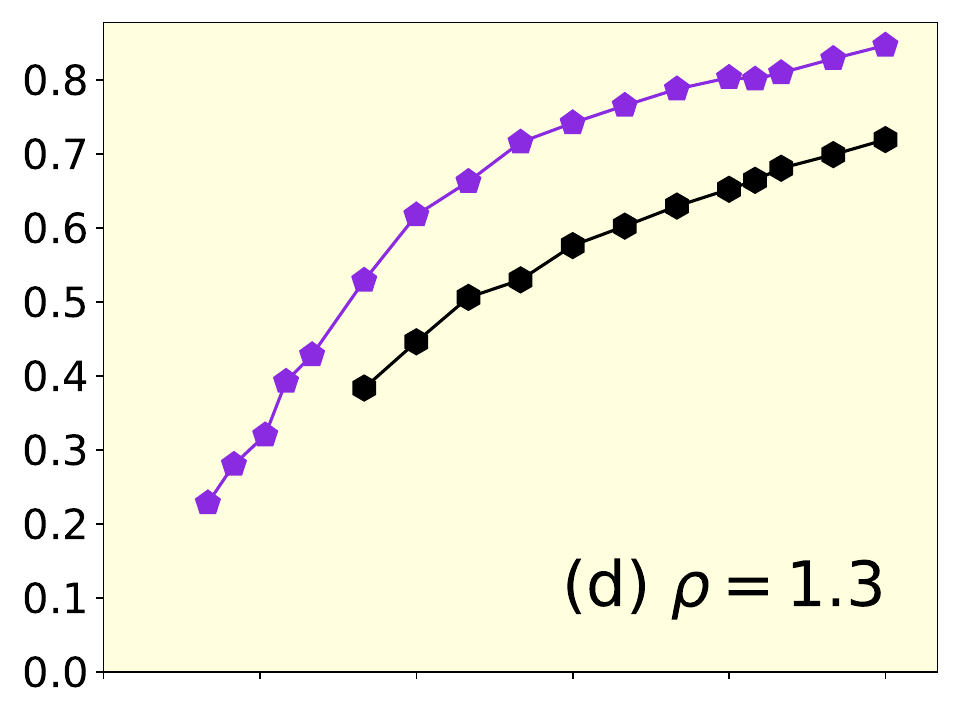, height=2.8cm,width = .23\textwidth, trim={0 0 0 0}}
\psfig{figure=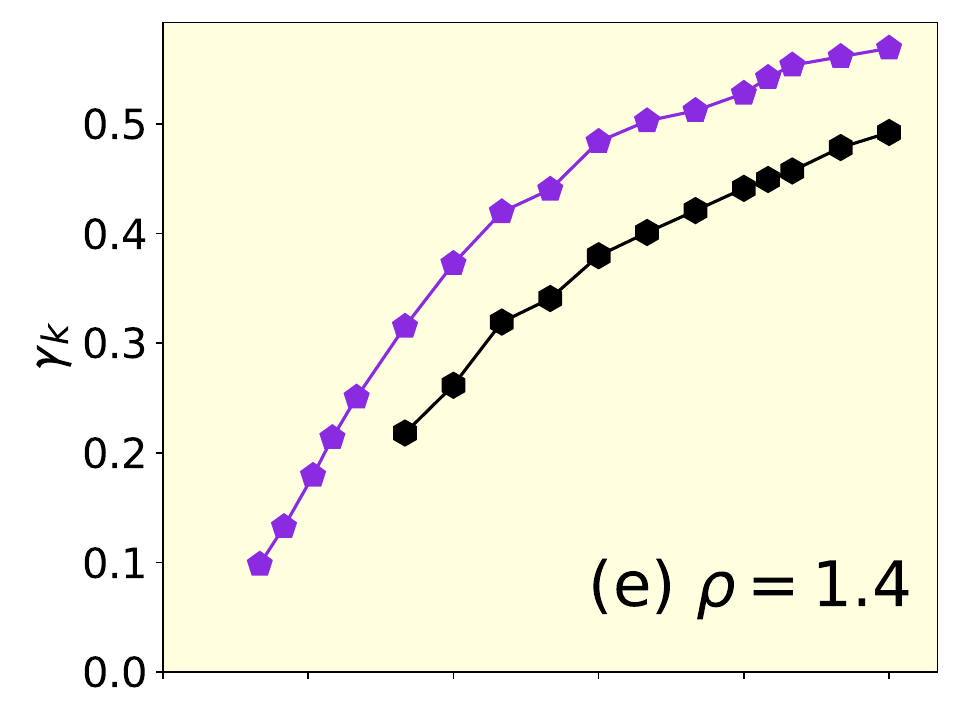, height=2.8cm,width = .23\textwidth, trim={0 0 0 0}}
\psfig{figure=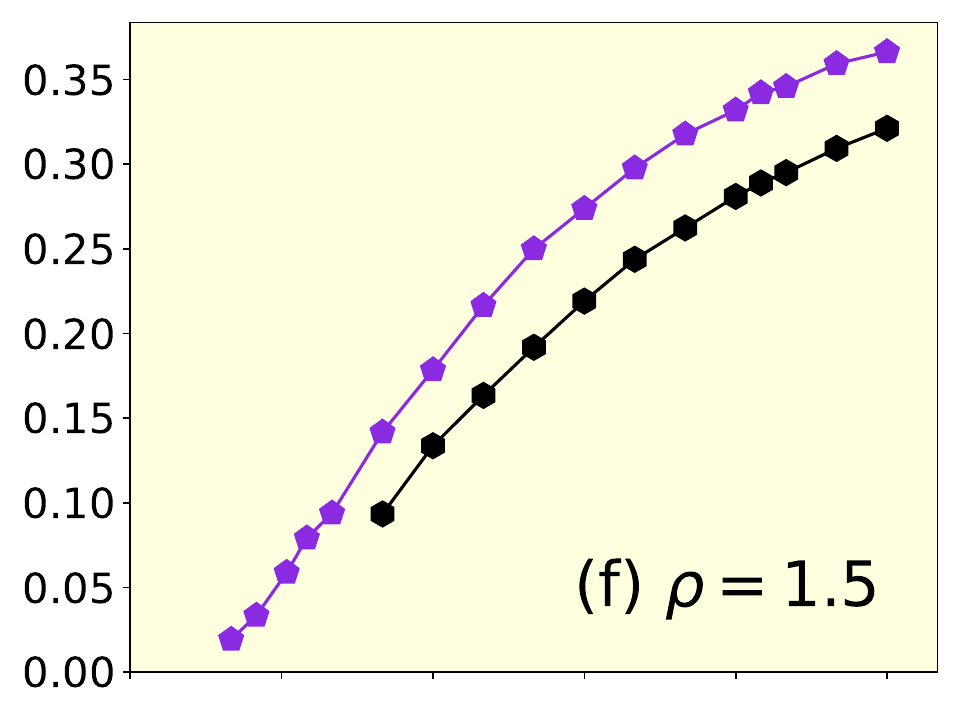, height=2.8cm,width = .23\textwidth, trim={13 0 0 0}}
\psfig{figure=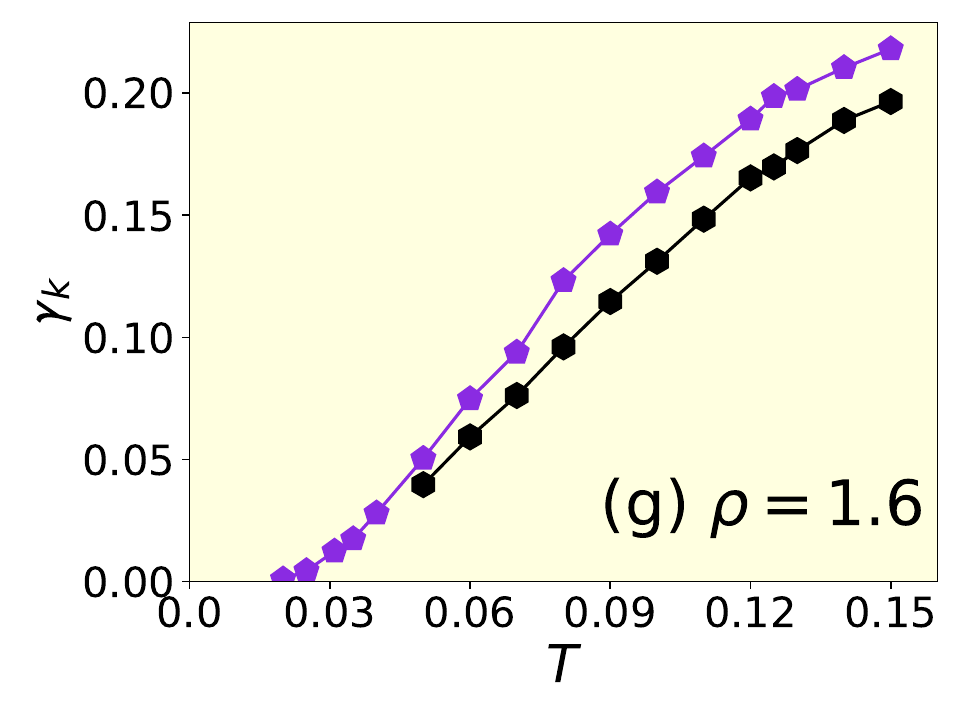, height=3.0cm,width = .23\textwidth, trim={13 0 0 0}}
\psfig{figure=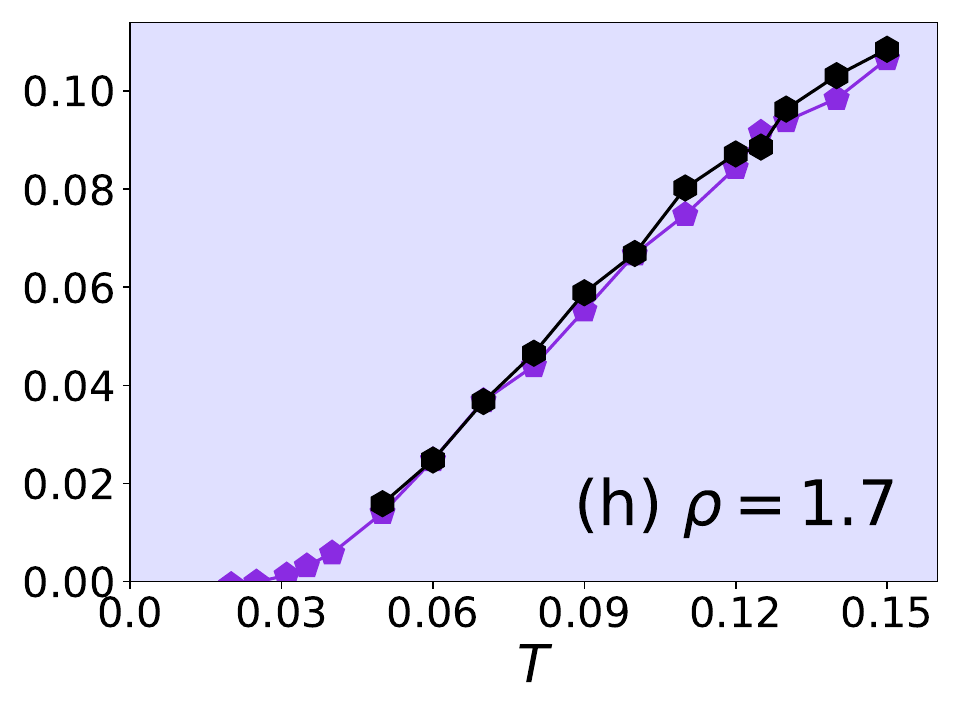, height=3.0cm,width = .23\textwidth, trim={13 0 0 0}}
\caption{Comparison of antinodal scattering rate $\gamma_k$ between $N_c=4, 16$ simulations for $\epsilon_p = 6.0$. The curves at lower temperature reveal additional features unavailable in large $N_c=16$ simulations.}
\label{Nc4}
\end{figure}

The intriguing downturn features motivates us to further explore the associated physics at lower temperatures.
Whereas the effectively stronger interaction effects for larger $\epsilon_p=6.0$ does not allow us to get access to lower temperatures, the isotropy of the scattering rate revealed in Fig~\ref{fig2} fortunately permits adopting smaller DCA cluster like $N_c=4$ to partly alleviate the QMC sign problem for simulating lower $T$ systems.

Figure~\ref{Nc4} compares the results of $\gamma_k$ at antinodal direction between $N_c=4$ and $N_c=16$, which generically display quantitative deviation except for high density e.g. $\rho=1.7$.
It can be seen that the dominant feature is the two consecutive nearly linear-$T$ regimes of antinodal $\gamma_k$ connected via a smooth crossover around $T\sim0.08$ for intermediate densities as shown in panels (c-g). Nonetheless, this feature is only obvious for $N_c=4$ simulations while larger $N_c=16$ seemingly smooths these out so that might question the physical reality of this phenomenology in the Emery model. Despite of this uncertainty, this two linear regimes are still significant in terms of its similarity with other lattice models e.g. single  band model in triangular lattice~\cite{triangular}.

Physically, this is reminiscent of the recent experimental demonstration of the crossovers between different temperature regimes signaling the Fermi liquid (FL), strange metal (SM), and empirical high-temperature bad metal state (similar to the strange metal but does not host well-defined quasiparticle)~\cite{xinghui}. 
In fact, the earlier theoretical calculations based on large-U Hubbard and t-J models have uncovered similar behavior~\cite{Shastry1,Shastry2,Shastry3,Shastry4}. At the lowest temperature, the resistivity is typically proportional to $T^2$ as a FL. Increasing temperature induces the linear resistivity indicating the SM regime. At even higher temperatures, the system enters into a bad metal regime with a different slope of the linear-$T$ resistivity.

Apart from the feature of two linear-$T$ regimes, the low density limit shows saturated scattering rate, which cannot be seen at large $N_c=16$ simulations. This again implies for the intrinsically strong interaction effects arising from large $\epsilon_p=6.0$. In contrast, at high density $\rho=1.7$, the perfect agreement between $N_c=4, 16$ at high temperatures provides evidence on the crossover to the apparent $T^2$ behavior seen for $N_c=4$ curve at lower temperatures. This $N_c=4$ simulations complements our understanding of ultimate fate of the linear-$T$ scattering rate with an unphysical negative interception extrapolated to $T=0$.

\begin{figure}
\psfig{figure=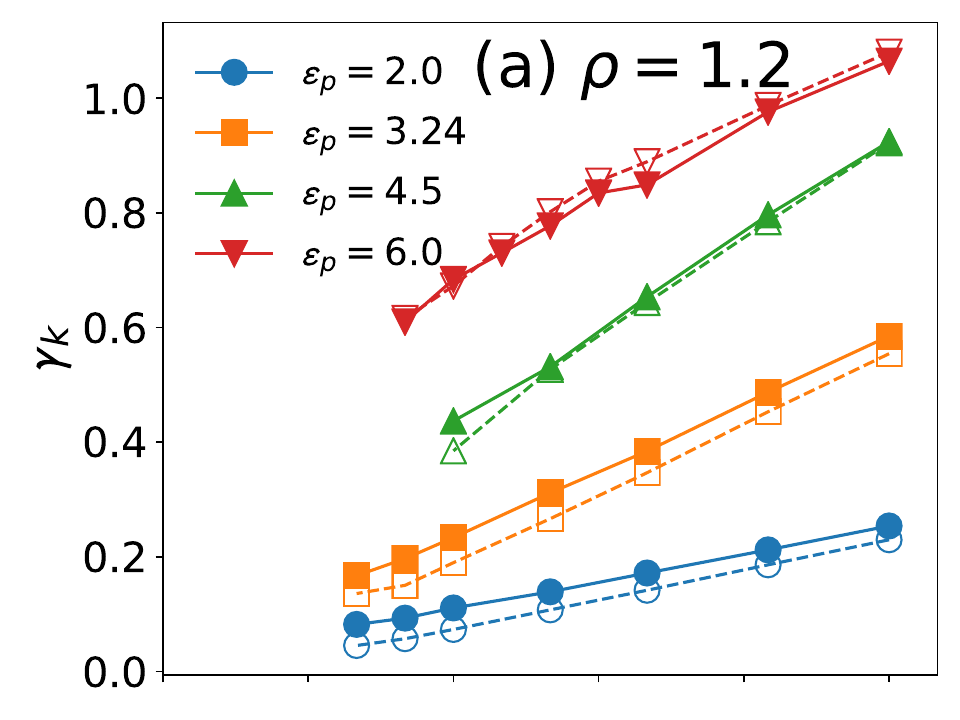, height=2.8cm,width = .23\textwidth, trim={0 0 0 0}}
\psfig{figure=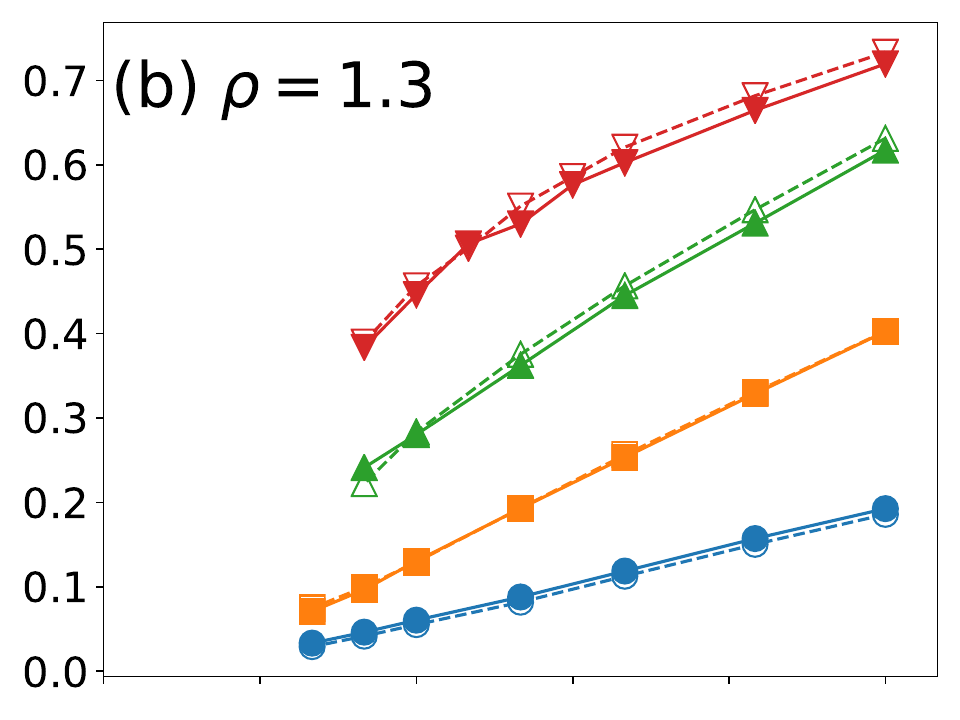, height=2.8cm,width = .23\textwidth, trim={0 0 0 0}}
\psfig{figure=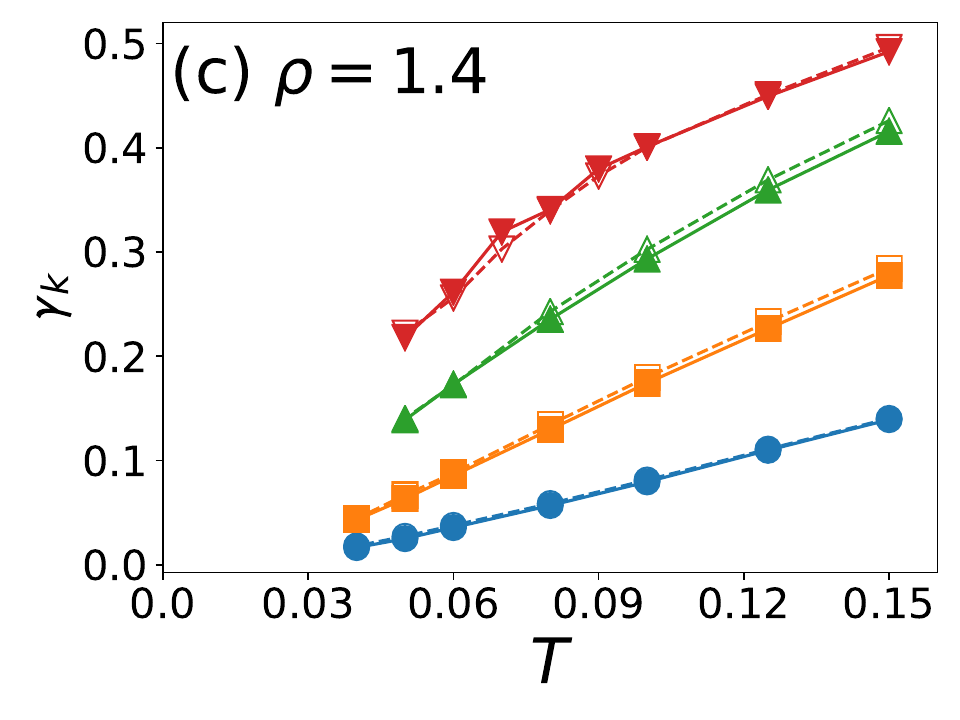, height=3.0cm,width = .23\textwidth, trim={0 0 0 0}}
\psfig{figure=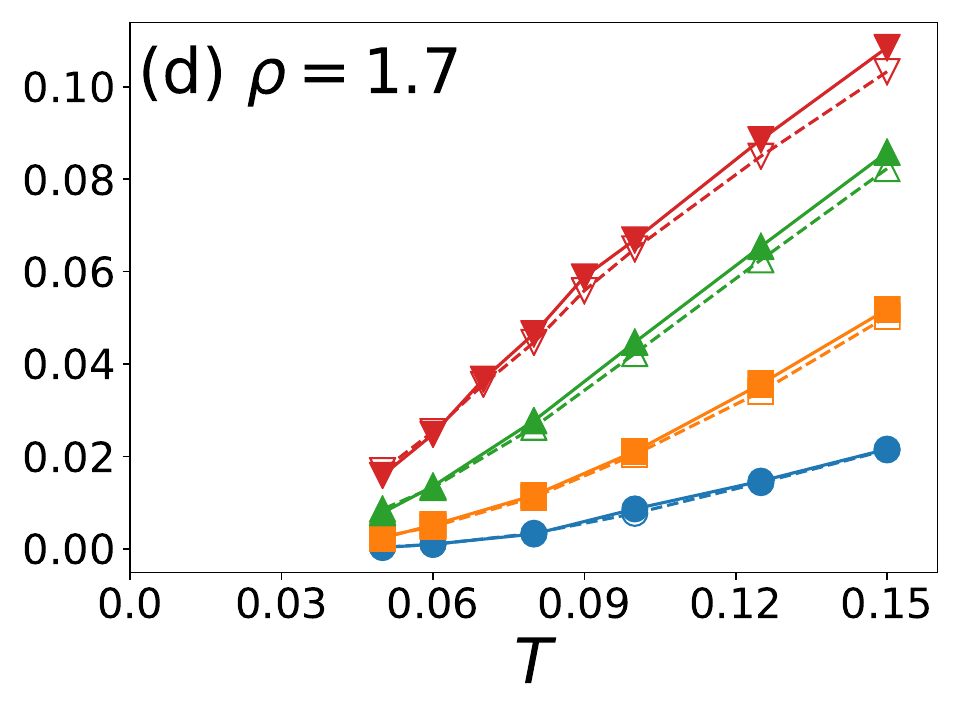, height=3.0cm,width = .23\textwidth, trim={12 0 0 0}}
\caption{$\epsilon_p$ dependence of the electronic scattering rate $\gamma_k$ at antinodal (solid) and nodal (dashed) directions, whose downturn occurs for relatively large $\epsilon_p=4.5, 6.0$ at intermediate densities of panels (b-c). The increase of $\epsilon_p$ generically leads to larger scattering rate reflecting stronger interaction effects.}
\label{ep}
\end{figure}

In addition to the two characteristic values of $\epsilon_p=$3.24 and 6.0, Figure~\ref{ep} provides more information on the $\epsilon_p$ dependence of the scattering rate. Here, the deviation between antinodal (solid) and nodal (dashed) $\gamma_k$ is mostly dominant at smaller density like $\rho=1.2$ while can be neglected at larger densities.
Apparently, the downturn of $\gamma_k$ occurs only for relatively large $\epsilon_p=4.5, 6.0$ at intermediate density e.g. $\rho=1.3, 1.4$ in panels (b-c). It is naturally expected that the downturn folding might be even more obvious at larger $\epsilon_p>6.0$ while the temperature scale is always around $T\sim 0.1$.

Undoubtedly, Fig.~\ref{ep} also illustrates that the increase of $\epsilon_p$ generically leads to higher scale of the scattering rate, which matches with the physical expectation that larger $\epsilon_p$ promotes higher carrier density on $d$ orbital so that effectively induce stronger interaction effects. Interestingly, if naively assuming the $T^{\alpha}$ dependence of $\gamma_k$, panel (d) at large density $\rho=1.7$ vividly displays the gradual crossover from the $\alpha>1$ at small $\epsilon_p=2.0$ to $\alpha<1$ (downturn) at large $\epsilon_p=6.0$.

\begin{figure*} 
\psfig{figure=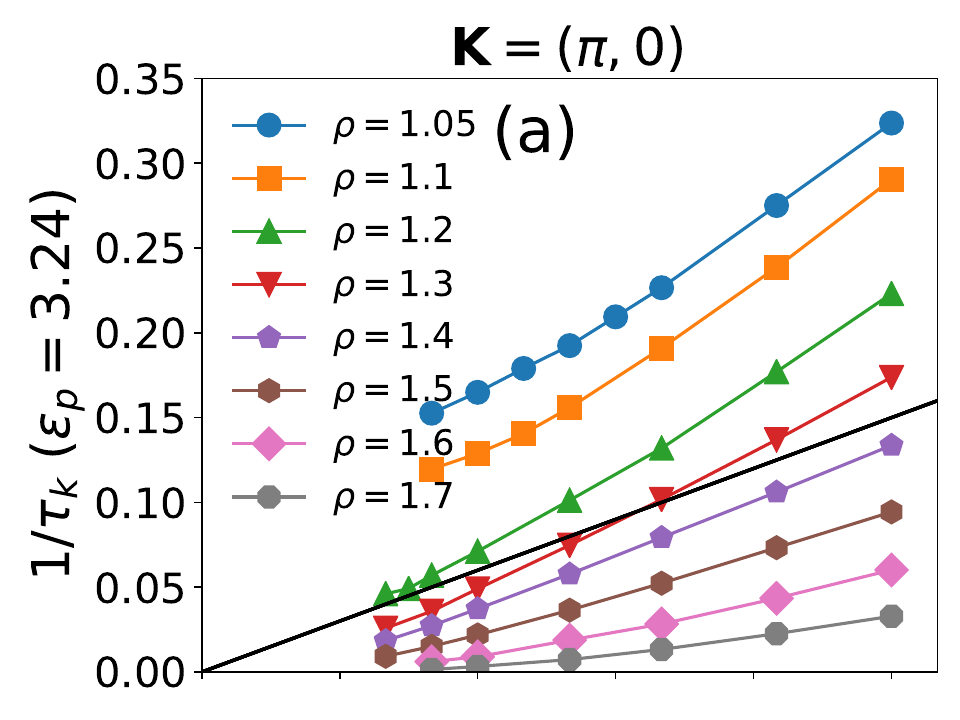, height=2.8cm,width = .23\textwidth, trim={0 10 0 0}}
\psfig{figure=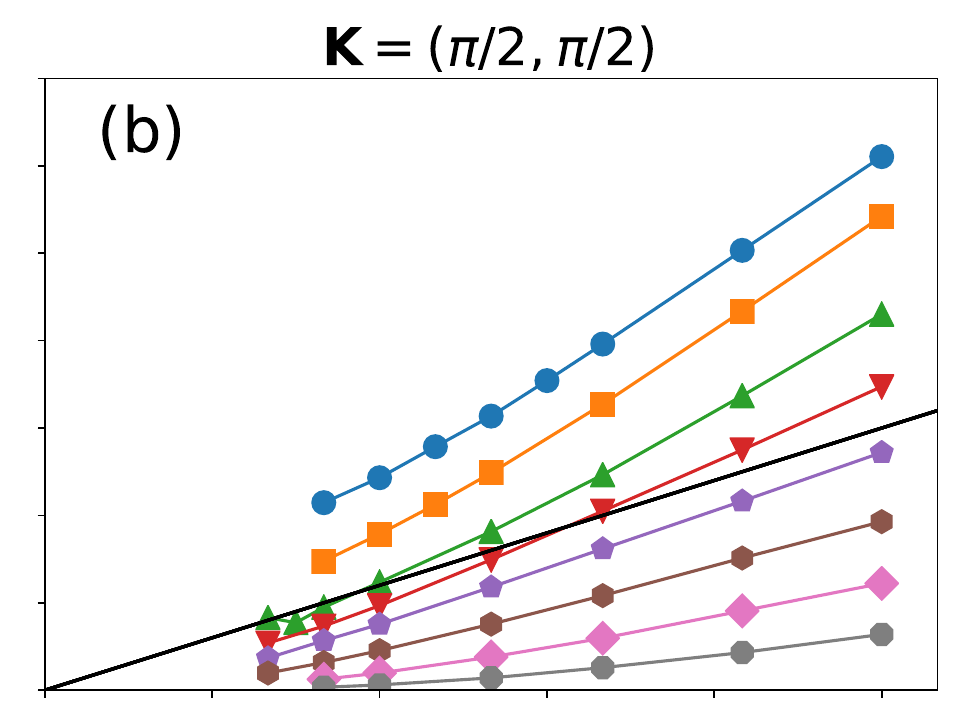, height=2.8cm,width = .19\textwidth, trim={0 0 0 0}}
\psfig{figure=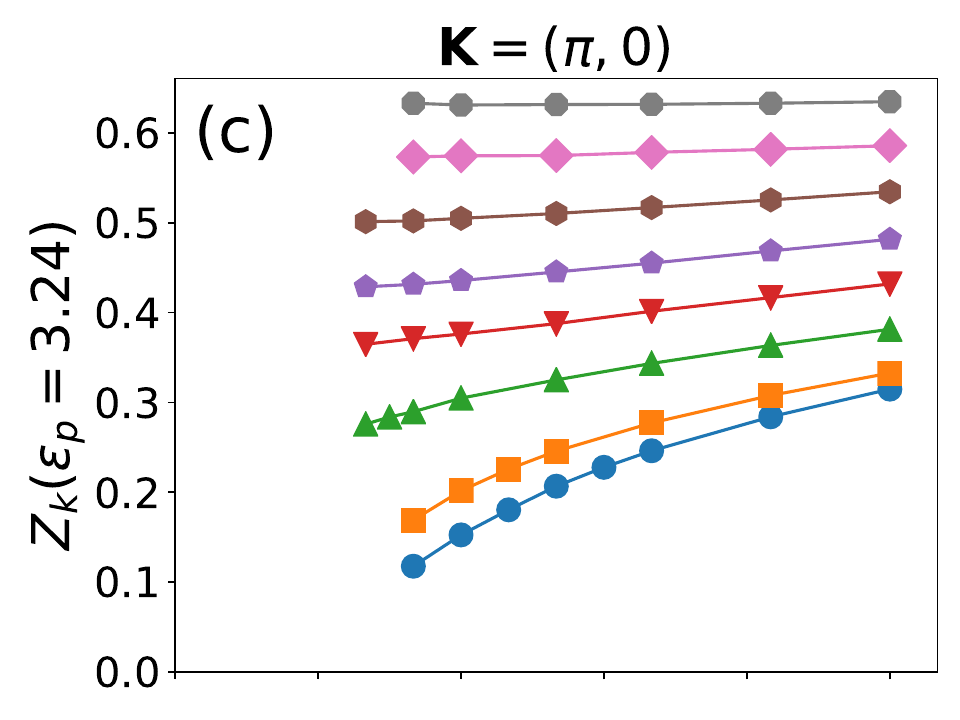, height=2.8cm,width = .23\textwidth, trim={0 10 0 0}}
\psfig{figure=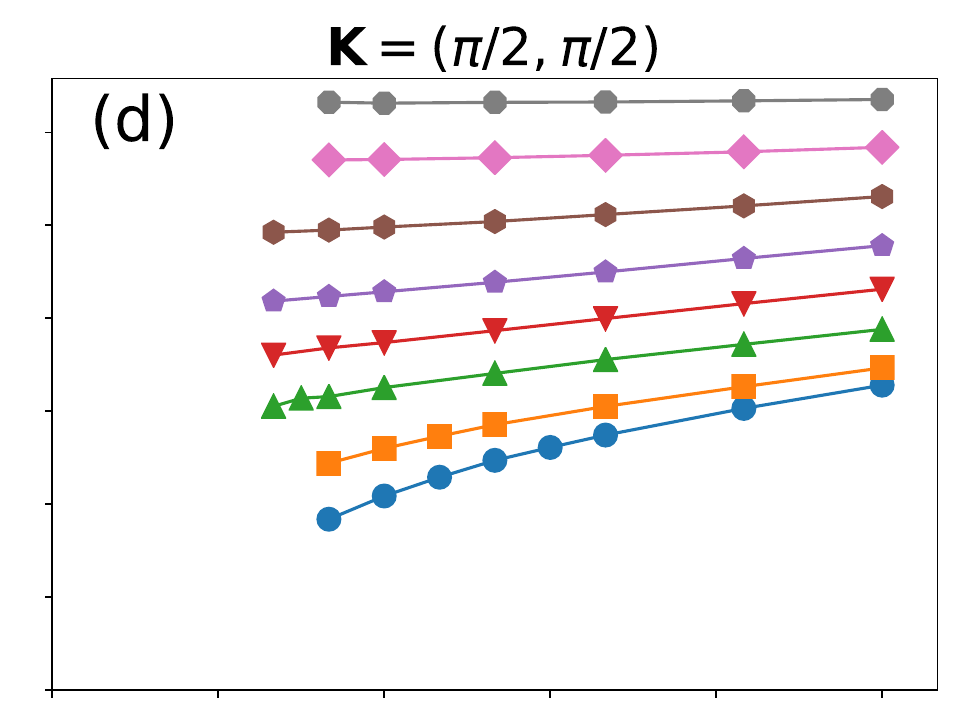, height=2.8cm,width = .19\textwidth, trim={0 0 0 0}}
\psfig{figure=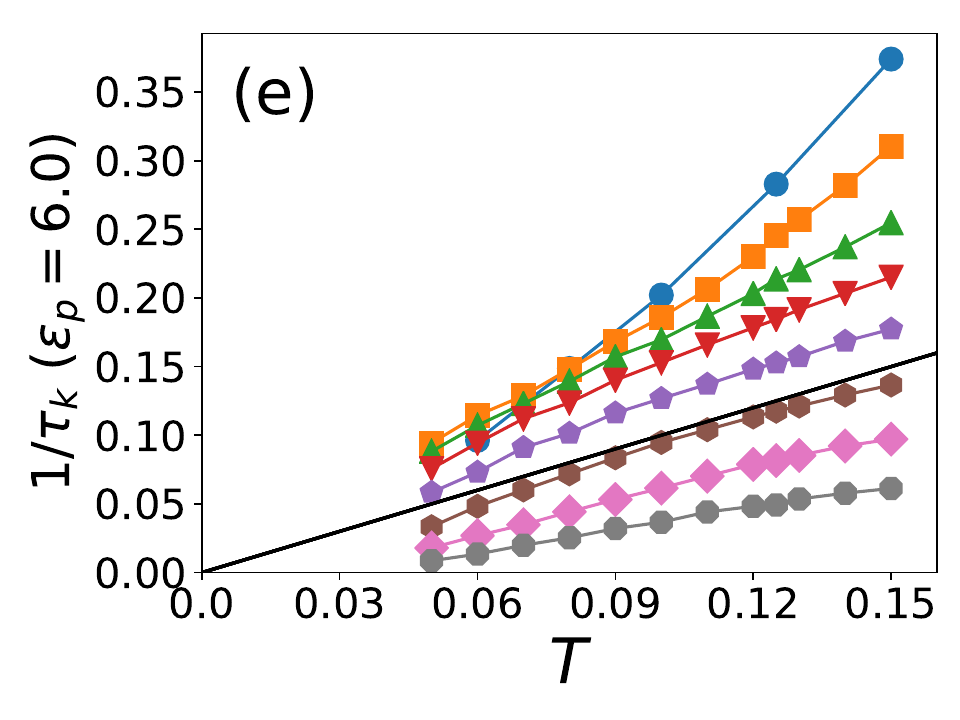, height=3.0cm,width = .23\textwidth, trim={0 0 0 0}}
\psfig{figure=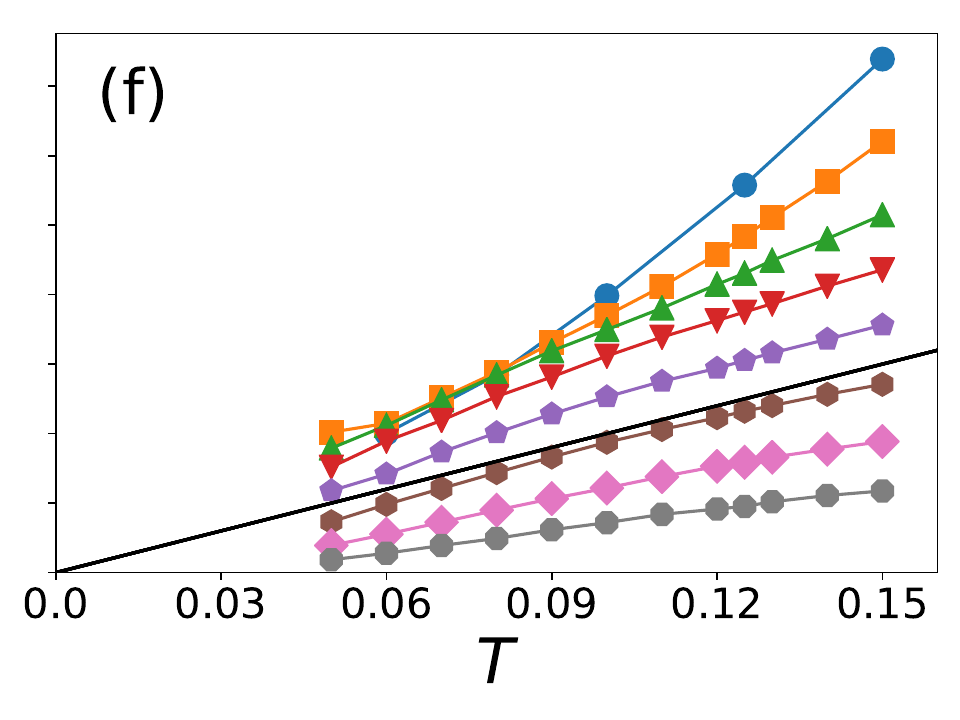, height=3.0cm,width = .19\textwidth, trim={5 0 0 0}}
\psfig{figure=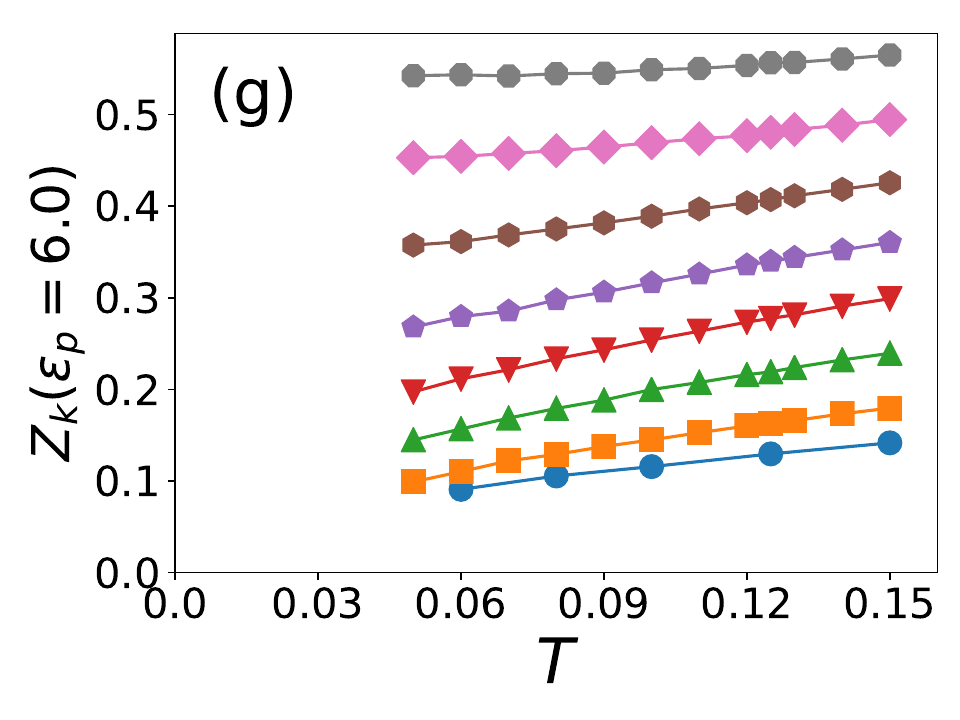, height=3.0cm,width = .23\textwidth, trim={0 0 0 0}}
\psfig{figure=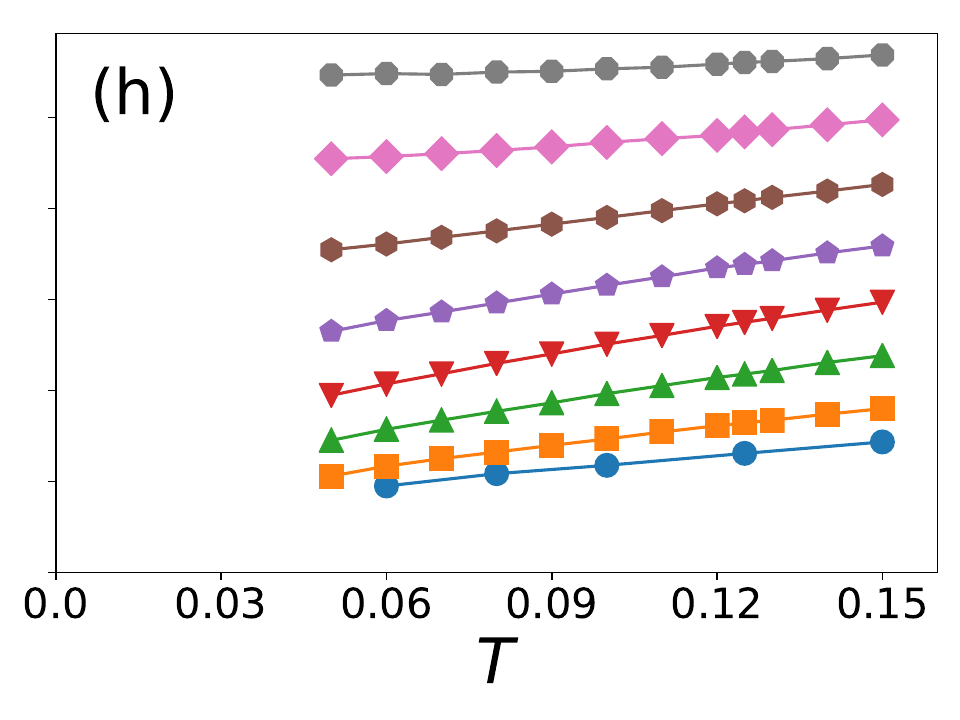, height=3.0cm,width = .19\textwidth, trim={5 0 0 0}}
\caption{Temperature dependence of the momentum-resolved quasiparticle scattering rate $1/\tau_k$ and quasiparticle weight $Z_k$ at nodal $(\pi/2,\pi/2)$ and antinodal $(\pi,0)$ directions for various density at (Upper) $\epsilon_p=3.24$ and (Lower) $\epsilon_p=6.0$.}
\label{tauk}
\end{figure*}

\subsection{Quasiparticle weight and scattering rate}

In the presence of electronic interaction, it is worthwhile further investigating the quasiparticle scattering rate or inverse quasiparticle life-time $1/\tau_k = Z_k \gamma_k$, which differs from the electronic scattering rate $\gamma_k$ by the so-called quasiparticle weight $Z_k$. The consideration of the quasipariticle scattering rate is essential to account for the many-body interaction effects within strongly correlated systems and is also more closely related to the experimental transport properties than the single particle electronic scattering rate $\gamma_k$ discussed earlier~\cite{Nolting}. 

Numerically, $Z_k$ can be approximately evaluated~\cite{triangular} by 
\begin{align}
    Z_k \approx [1-\frac{\Im\Sigma(i\omega_0)}{\omega_0}]^{-1}
\end{align}

One prediction of the Planckian dissipation theory~\cite{Zaanen04,Rev3,Zaanen19} is that the inverse quasiparticle lifetime $1/\tau_k$ is proportional to absolute temperature T with a coefficient close to unity~\cite{Zaanen04,Rev3,Zaanen19}, which is found to be quantitatively incompatible with the previous numerical studies on single band model on square~\cite{nfl_pnas} and triangular lattice~\cite{triangular}.

Figure~\ref{tauk} demonstrates that this prediction of unity slope only crudely holds true for $\rho \sim 1.4$ at $\epsilon_p=3.24$ and for $\rho \sim 1.5$ at $\epsilon_p=6.0$ (note that black dashed line has unity slope for reference). In this sense, we have not found any decisive signature for the unity slope feature associated with the universal Planckian limit.
Apart from the slope, the general trend of the quasiparticle scattering rate in terms of the density is similar to the electronic scattering rate $\gamma_k$ for both nodal and antinodal directions. Besides, the bottom panels show that there also exists the downturn of $1/\tau_k$ at temperatures below $T\sim0.1$ for intermediate densities, which matches with the behavior of $\gamma_k$ generically.

Fig.~\ref{tauk} also displays the quasiparticle weight $Z_k$ that is capable to characterize its closeness to conventional Fermi liquid (with $Z_k=1$).
Undoubtedly, $Z_k$ increases with the density reflecting the role of doped holes. Additionally, it is always weakly temperature dependent or even independent except for the low doping $\rho=1.05, 1.1$ at $\epsilon_p=3.24$, which is normally associated with their PG features.

\section{Summary and outlook}

In summary, motivated by the recent numerical exploration of the non-Fermi liquid signatures in the single-band Hubbard model on square~\cite{nfl_pnas} and triangular lattices~\cite{triangular}, we have employed the dynamic cluster quantum Monte Carlo calculations to systematically investigate the temperature dependence of the electronic and quasiparticle scattering rates in the framework of two dimensional three-orbital Emery model, which is generically believed to capture the physics of cuprate SC more accurately~\cite{3b_dqmc,3b_num,nfl_pnas}.

Our numerical simulations reveal that the systems with moderate site energy of $p$-orbital $\epsilon_p=3.24$ relevant to cuprates support the linear-in-$T$ scattering rates for a range of intermediate densities $\rho=1.2-1.5$, while the small densities feature PG behavior and large densities show the conventional Fermi liquid characterized by $T^2$ scattering rates. Nonetheless, in many cases, the interception at $T=0$ is negative indicating that other factors have to be included to account for the physical reality.

Our study also extends to the relatively large  $\epsilon_p=6.0$ presumably relevant to newly discovered nickelate SC. For these systems, the common feature lies in the downturn of the scattering rate below the temperature scale $T\sim0.1$. Our simulations at smaller DCA cluster $N_c=4$, which is reasonable owing to the isotropy of the scattering rate, confirms this observation by revealing two consecutive linear-$T$ regimes of $\gamma_k$. More simulations on other $\epsilon_p$ values provide further evidence that only relatively large $\epsilon_p$ would induce such a downturn of $\gamma_k$. 
To have a complete understanding of the many-body effects, we also illustrated the behavior of quasiparticle scattering rate, which generically departs from the unity slope as the prediction of Planckian dissipation theory. The quasiparticle weight is monotonically increasing with the hole density as expected owing to the enhanced metallicity from the doped charge carriers.

Overall, our numerical findings of the three-orbital Emery model generally match with those observed in two-dimensional Hubbard model~\cite{nfl_pnas}. The hole doping/density range of NFL state is very narrow. Hindered by the negative sign problem, however, we are unable to definitively identify the doping range exhibiting perfect linear-in-temperature behavior in large enough DCA cluster. 
Our presented work provides quantitative examination of linear-$T$ features of the scattering rates in the celebrated Emery model. 

Because our current investigation only focused on the scattering rate but neglected other closely related physics such as SC and magnetic/charge ordering, some further directions deserve future exploration. For instance, the most recent transport experiments have revealed some relations between the superconducting $T_c$ and the strange-metal's slope $A$ as $T_c \sim \sqrt{A}$ in the cuprate SC~\cite{jinkui}. The detailed studies on the connection between SC and the behavior of scattering rate would be interesting. Besides, despite that relatively large $\epsilon_p$ might be detrimental to SC~\cite{3b_num}, it can be fruitful to find another physical quantities having decisive relation with the behavior associated with the scattering rate uncovered in the present work. Another direction might be more investigation on the high hole or electron doping systems since our current study at large hole density $\rho=1.7$ has demonstrated interesting linear-$T$ behavior of the scattering rate.

In addition, in light of the most recent experimental demonstration of the cuprate-like electronic structure of infinite-layer nickelates~\cite{ding2024cupratelike,sun2024electronic} revealing the dominant role of $d_{x^2-y^2}$ orbital and the previous evidence of the large charge transfer energy in these nickelates, our current investigation of the Emery model at relatively large $\epsilon_p$ would provide particularly important information on the nickelates. 
This exploration sheds light on some fundamental factors governing the physics of nickelates and cuprates.
However, it is still questionable that whether the uncovered three-dimensional (3D) electron pocket centered at
Brillouin zone corner originating from the rare-earth atoms of the nickelates plays the vital or only marginal roles.

\section{Acknowledgement}
We would like to thank Wenxin Ding, Wei Wu, and Peizhi Mai for illuminating discussions in the early stage.
We acknowledge the support by National Natural Science Foundation of China (NSFC) Grant No.~12174278, startup fund from Soochow University, and Priority Academic Program Development (PAPD) of Jiangsu Higher Education Institutions.

\bibliographystyle{apsrev4-1}
\bibliography{main}

\end{document}